\documentclass[twocolumn]{aastex62}

\usepackage{amsmath}
\usepackage{hyperref}

\newcommand{\integral}{\textit{INTEGRAL}}
\newcommand{\nustar}{\textit{NuSTAR}}
\newcommand{\xmm}{\textit{XMM-Newton}}
\newcommand{\chandra}{\textit{Chandra}}
\newcommand{\swift}{\textit{Swift}} 
\newcommand{\fermi}{\textit{Fermi}}
\newcommand{\msol}{\,M$_\odot$}
\newcommand{\ergcms}{\,erg\,cm$^{-2}$\,s$^{-1}$}
\newcommand{\phcms}{\,ph\,cm$^{-2}$\,s$^{-1}$}

\newcommand{\nh}{$N_{\rm H}$}
\newcommand{\cm}{\,cm$^{-2}$}

\usepackage{ulem}

\received{March 8, 2019} %Month Day, 2018
\revised{September 12, 2019} %\today
\accepted{October 5, 2019}
\shorttitle{\textit{NuSTAR} observations of unidentified IGR sources}
\shortauthors{Clavel et al.}
\begin{document}

\title{\textit{NuSTAR} observations of the unidentified \textit{INTEGRAL} sources: constraints on the Galactic population of HMXBs}

\correspondingauthor{Ma\"ica Clavel}
\email{maica.clavel@univ-grenoble-alpes.fr}

\author[0000-0003-0724-2742]{Ma\"ica Clavel}
\affil{Space Sciences Laboratory, 7 Gauss Way, University of California, Berkeley, CA 94720-7450, USA}
\affil{Univ. Grenoble Alpes, CNRS, IPAG, F-38000 Grenoble, France}

\author[0000-0001-5506-9855]{John A. Tomsick}
\affiliation{Space Sciences Laboratory, 7 Gauss Way, University of California, Berkeley, CA 94720-7450, USA}

\author[0000-0002-8548-482X]{Jeremy Hare}
\affiliation{Space Sciences Laboratory, 7 Gauss Way, University of California, Berkeley, CA 94720-7450, USA}

\author[0000-0003-2737-5673]{Roman Krivonos}
\affiliation{Space Research Institute of the Russian Academy of Sciences, Profsoyuznaya Str. 84/32, 117997 Moscow, Russia}

\author[0000-0002-9709-5389]{Kaya Mori}
\affiliation{Columbia Astrophysics Laboratory, Columbia University, New York, NY 10027, USA}

\author[0000-0003-2686-9241]{Daniel Stern} 
\affiliation{Jet Propulsion Laboratory, California Institute of Technology, Pasadena, CA 91109, USA}



\begin{abstract}

The \nustar\ Legacy program `Unidentified \integral\ sources' targeted faint hard X-ray sources revealed by \integral\ in the Galactic plane in order to provide conclusive identification of their nature and insights on the population of faint hard X-ray sources. The \nustar\ and \swift/XRT observations obtained in 2015--2017 contributed to the successful identification of five persistent sources. Here, we report on the spectral and variability analyses which helped to consolidate the classifications of IGR\,J10447--6027, IGR\,J16181--5407 and IGR\,J20569+4940 as active galactic nuclei, and IGR\,J17402--3656 as an intermediate polar. An optical spectrum of the blazar IGR\,J20569+4940 is also presented. Combining these results with successful identifications of other such faint and persistent \integral\ sources reported in the literature, we investigate possible implications for the population of persistent high mass X-ray binaries (HMXBs) below the identification completion limit of the \integral\ survey. The current trend hints at a deficit of persistent HMXBs below $F_{\rm 17-60keV} = 10^{-11}$\,erg\,cm$^{-2}$\,s$^{-1}$, but additional efforts dedicated to classifying faint hard X-ray sources are needed before we can draw solid conclusions.

\end{abstract}



%%%%%%%%%% Section 1: Introduction
%%%%%%%%%%%%%%%%%%%%%%%%%%%%%%%%%%%%%%%%%%%%%%%%%%%%%%%%%%%%%%%%%%%%
%%%%%%%%%%%%%%%%%%%%%%%%%%%%%%%%%%%%%%%%%%%%%%%%%%%%%%%%%%%%%%%%%%%%
%%%%%%%%%%%%%%%%%%%%%%%%%%%%%%%%%%%%%%%%%%%%%%%%%%%%%%%%%%%%%%%%%%%%
%%%%%%%%%%%%%%%%%%%%%%%%%%%%%%%%%%%%%%%%%%%%%%%%%%%%%%%%%%%%%%%%%%%%
\section{Introduction}

Since its launch in 2002, the \integral\ mission \citep{winkler2003} has been surveying the hard X-ray sky. It has now revealed more than a thousand sources \citep[see e.g.][for the latest catalog and recent updates on deep surveys]{bird2016, mereminskiy2016, krivonos2017}, with about half of them in the direction of the Galactic plane \citep[$\vert b \vert < 17.5^\circ$;][]{krivonos2012}. 
About two-thirds of the objects detected in the plane of the Galaxy are indeed Galactic hard X-ray sources, mainly low-mass X-ray binaries (LMXBs), high-mass X-ray binaries (HMXBs) and cataclysmic variables (CVs).
 
Using the \integral\ source catalog established by \citet{krivonos2012}, \citet{lutovinov2013} studied the population of persistent HMXBs in the Galaxy ($\vert b \vert < 5^\circ$). Most of them ($\sim$80\%) host neutron stars (NS) which are accreting matter from the stellar wind of their massive companion. However, the identification completeness of this sample is limited to $F_{\rm 17-60keV} > 1\times 10^{-11}$\,erg\,cm$^{-2}$\,s$^{-1}$ with a large majority of the known persistent HMXBs above this limit (47 out of 53 in their study).
If these bright HMXBs are likely to have significant feedback onto the Galactic environment, and are expected to play an important role in the formation and the evolution of galaxies through cosmic times \citep[e.g.][]{brorby2016}, the putative presence of a larger population of fainter HMXBs should not be neglected.
Such low-luminosity HMXBs, characterized by a persistent X-ray behavior (or, alternatively, a very long outburst recurrence period) could be among the numerous unclassified faint hard X-ray sources, such as Galactic sources detected in the \nustar\ serendipitous survey \citep{tomsick2017} and faint \integral\ sources in the Galactic plane \citep[see][for a  recent conclusive identification of a transient HMXB]{hare2019}.

Below the identification completion limit of their sample, \citet{lutovinov2013} reported six persistent HMXBs and twenty-six unidentified sources. 
Extrapolating the luminosity function derived for the bright persistent HMXBs, they predicted the existence of at least twelve HMXBs with $F_{\rm 17-60keV} < 1\times 10^{-11}$\ergcms\ among persistent sources detected by \integral\  \citep[in the nine-year Galactic hard X-ray survey,][]{krivonos2012}, i.e.\ at least six among the twenty-six sources that were unidentified at the time.  
However, if the luminosity function derived by \citet{lutovinov2013} is extrapolated towards the fainter fluxes with no major modification in their properties, the distribution of HMXBs could also present a gap that would for instance highlight a modification in the accretion mechanisms \citep[e.g.\ direct accretion versus `propeller state',][]{postnov2017}, but it could also flatten, with none or very few objects present at low luminosities.
To test these predictions and better understand the likely progenitors of double degenerate binaries, whose mergers are responsible for the gravitational waves that have been detected \citep{abbott2018}, it is essential to characterize the population of faint hard X-ray sources in the Galaxy by improving the completeness of the current sample. 

\begin{figure*}
	\centering
	\includegraphics[width=0.9\textwidth]{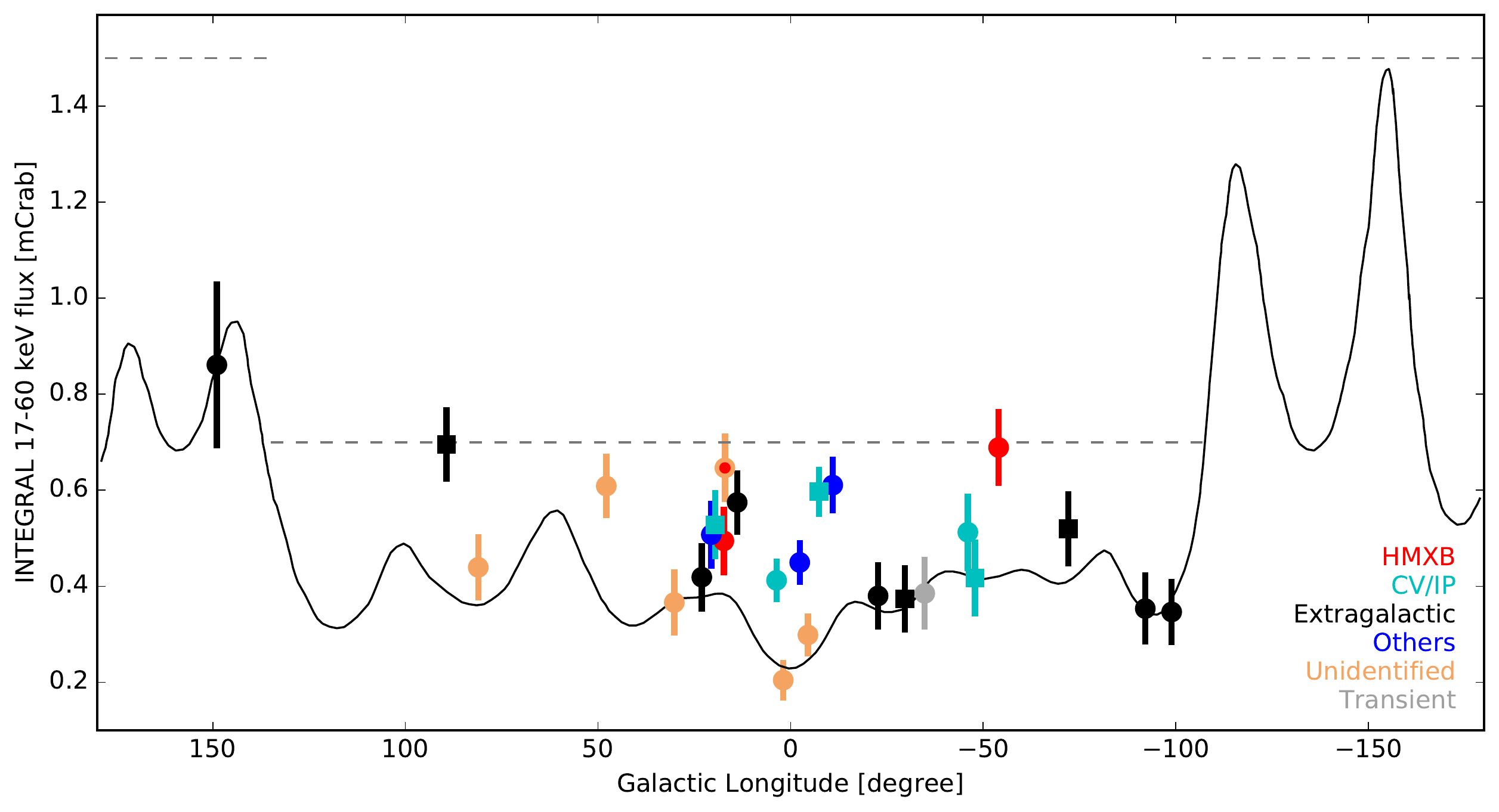}
	\caption{Distribution of persistent IGR sources listed as unidentified in \citet[][see also Table~\ref{tab:cat}]{lutovinov2013}. Their 17--60\,keV flux is below the limit chosen by \citet{lutovinov2013} to investigate the persistent HMXB population (dashed lines) and close to the \integral\ sensitivity limit at 4.7\,$\sigma$ in the 9-year catalog \citep[solid line,][]{krivonos2012}. The orange color shows the sources that remain unidentified, while the others have a color code highlighting their nature (black: extragalactic sources, cyan: intermediate polars, red: high-mass X-ray binaries, blue: other types). XTE J1824--141 is both orange and red because its identification as a pulsar makes it more likely to be an HMXB than the five truly unidentified sources, for which we only have very limited information. Squares mark the six sources that have been observed by \nustar. 
	}
	\label{fig:distrib}
\end{figure*}

The ‘unidentified \integral\ sources’ Legacy program conducted by the \textit{Nuclear Spectroscopic Telescope Array} \citep[\nustar,][]{harrison2013} takes full advantage of \nustar's better sensitivity and higher spatial resolution to investigate the faint persistent sources detected by \integral. The present classification status regarding the twenty-six unidentified sources listed by \citet{lutovinov2013} is summarized in Table~\ref{tab:cat} and Figure~\ref{fig:distrib}. Twenty sources have now been identified, including six sources that have been observed by \nustar. 
Five of these were observed as part of the Legacy program with results from IGR\,J18293--1213 being reported in \citet{clavel2016} and results from the other four being reported in this work.  In addition, the \nustar\ results for IGR\,J14091--6108 were reported in \citet{tomsick2016b}.

\begin{table*} 
        \centering
        \caption{Persistent IGR sources listed as unidentified in \citet{lutovinov2013}. The first four columns, i.e., the source names, their Galactic coordinates and their hard X-ray fluxes, are adopted from the \integral\ catalog \citep{krivonos2012}. The last two columns give the source identifications (as in 2019) and the most-recent references. The `\nustar' mention highlights the sources that have been observed by \nustar, including the four sources presented in Sect.~\ref{sec:nustar} (`This work'). The list is separated into six groups, corresponding to the color coding used in Figure~\ref{fig:distrib}.
        }
        \label{tab:cat}
		\begin{tabular}{l r r c c c c c}
				\hline \hline
				Name &  \multicolumn{1}{c}{$l$} &  \multicolumn{1}{c}{$b$} & 17--60keV Flux &   Type  & Reference\\
				          & (deg) & (deg) & (mCrab)$^{*}$ &    (2019)  &\\
				\hline
				IGR J04059+5416      & 148.93 & 1.54 & 0.8611 &     AGN &  \citet{tomsick2015}\\
				IGR J08297--4250      & --98.92 & --2.21 & 0.3465 &    AGN & \citet{tomsick2015}\\ 
				IGR J09189--4418   &  --92.07   &  3.65  & 0.3535 &   AGN & \citet{tomsick2012}\\ 
				IGR J10447--6027      & --72.10 & --1.32 & 0.5196 &    AGN & \nustar\ - This work, \citet{fortin2018}\\
				IGR J16181--5407      & --29.66 & --2.63 & 0.3736 &        AGN & \nustar\  - This work\\
				IGR J16560--4958  &   --22.68 &     --4.17 &    0.3800 &  AGN  & \citet{tomsick2012} \\ 
				IGR J18134--1636      & 13.87 & 0.58 & 0.5746 &      AGN &  \citet{zolotukhin2015}\\ 
				IGR J18381--0924      & 23.05 & --1.36 & 0.4189 &    AGN &  \citet{rahoui2017} \\ 
				IGR J20569+4940      & 89.31 & 2.76 & 0.6952 &                AGN & \nustar\ \& Keck/LRIS - This work\\
				\hline
				IGR J14091--6108      & --47.84 & 0.33 & 0.4175 &    CV/IP & \nustar\ - \citet{tomsick2016b}\\
				IGR J14257--6117   &  --45.99 &    --0.46 &   0.5125 &  CV/IP &  \citet{bernardini2018} \\
				IGR J17402--3656      & --7.38 & --3.27 & 0.5970 &   CV/IP & \nustar\ - This work, \citet{fortin2018} \\
				IGR J18088--2741      & 3.65 & --3.84 & 0.4124 &    CV/IP &  \citet{rahoui2017}\\
				IGR J18293--1213      & 19.56 & --0.71 & 0.5282 &    CV/IP & \nustar\ -  \citet{clavel2016} \\
				\hline
				IGR J13186--6257     & --53.99 &   --0.24 &   0.6890 &  HMXB & \citet{dai2011}\\ 
				IGR J18219--1347  &    17.33 &   0.12 & 0.4945 &   HMXB &  \citet{laparola2013}\\ 
				\hline
				IGR J17164--3803      & --10.94 & 0.07 & 0.6107 &    Symbiotic star &  \citet{rahoui2017}\\ 
				IGR J17233--2837      & --2.37 & 4.26 & 0.4496 &    LMXB/Pulsar &  \citet{bogdanov2014} \\
				IGR J18256--1035  &    20.60  &   0.80 &  0.5076 &  LMXB? &  \citet{masetti2013a}\\ 
				\hline
				IGR J17315--3221      & --4.50 & 0.81 & 0.2986 & ? &   --- \\
				AX J1753.5--2745      & 1.91 & --0.88 & 0.2046 & ? &   ---   \\ 
				XTE J1824--141 $^{**}$   &   17.04  & --0.71 &  0.6465 &  HMXB?/Pulsar & \citet{Markwardt2008}\\			
				IGR J18497--0248      & 30.19 & --0.93 & 0.3662 & ?   & --- \\
				IGR J19113+1413      & 47.84 & 2.13 & 0.6088 & ? &   --- \\
				Swift J2037.2+4151   & 81.08 & 0.53 & 0.4394 & ? & ---   \\
				\hline
				IGR J15335--5420  &   --34.80 &    1.37 &   0.3853 &  Transient & \citet{tomsick2016a}\\	
				\hline \hline
		\end{tabular}
        \begin{flushleft}
        $^{*}$ In the 17--60\,keV range, $1$\,mCrab corresponds to $1.43\times10^{-11}$\,erg\,cm$^{-2}$\,s$^{-1}$.\\
        $^{**}$ XTE J1824--141 has been successfully identified as a pulsar, but the nature of its companion (high or low mass) is uncertain.
		\end{flushleft}
\end{table*}

We describe the \nustar\ and \swift/XRT observations obtained through the \nustar\ Legacy program in Section~\ref{sec:data}, followed by the corresponding data reduction and the spectral analysis we performed. In Section~\ref{sec:nustar}, we present the results obtained for four of the faint \integral\ sources we observed with \nustar, namely IGR\,J10447--6027 (AGN, Sect.~\ref{sec:J10447}), IGR\,J16181--5407 (AGN, Sect.~\ref{sec:J16181}), IGR\,J17402--3656 (CV/IP, Sect.~\ref{sec:J17402}) and IGR\,J20569+4940 (blazar, Sect.~\ref{sec:J20569}). In Section~\ref{sec:synthesis}, we use these identifications together with the fourteen classifications obtained by independent efforts to investigate the population of persistent high-mass X-ray binaries in our Galaxy. In particular, by making a realistic assumption about the number of HMXBs among the six remaining unidentified sources, we revise the surface density of these sources down to the detection limit of the \integral\ survey. We conclude in Section~\ref{sec:conclu}.

%%%%%%%%%%%%%%%%%%%%%%%%%%%%%%%%%%%%%%%%%%%%%%%%%%%%%%%%%%%%%%%%%%%%
%%%%%%%%%%%%%%%%%%%%%%%%%%%%%%%%%%%%%%%%%%%%%%%%%%%%%%%%%%%%%%%%%%%%
%%%%%%%%%%%%%%%%%%%%%%%%%%%%%%%%%%%%%%%%%%%%%%%%%%%%%%%%%%%%%%%%%%%%
%%%%%%%%%%%%%%%%%%%%%%%%%%%%%%%%%%%%%%%%%%%%%%%%%%%%%%%%%%%%%%%%%%%%
%\clearpage
\section{Data reduction and analysis}
\label{sec:data}
The Legacy program led by our group provided \nustar\ observations of five unidentified IGR sources along with short \swift/XRT exposures (see Table~\ref{tab:obs}).

\begin{table*} 
        \centering
        \caption{\nustar\ and \swift/XRT observations obtained through the `Unidentified IGR sources' \nustar\ Legacy program$^*$.}
        \label{tab:obs}
		\begin{tabular}{l r r c c c c c}
				\hline \hline
				Source & \multicolumn{1}{c}{R.A.} & \multicolumn{1}{c}{Dec.} & Observatory & Obs. ID & Date & Exposure & Reference\\
				& \multicolumn{2}{c}{(J2000, deg)$^{**}$}  & & & & (ks) & \\				
			
				\hline
				
				IGR J10447--6027  & 161.21621 & --60.42000   & \nustar & 30161003002 & 2015-09-25 & 11.9 & This work \\
						&&& \swift/XRT & 00081765001,2 && 1.9 &\\
						&&& \nustar & 30161003004 & 2015-11-27 & 42.2 &\\
						&&& \swift/XRT & 00081765003 && 2.0 &\\
				IGR J16181--5407  & 244.53342 & --54.10272    & \nustar & 30161006002 & 2017-01-29 & 52.0 & This work\\
						&&& \swift/XRT & 00081977001 && 1.4 &\\
				IGR J17402--3656  & 265.11192 & --36.92706 & \nustar & 30161004002 & 2016-04-26 & 26.0 & This work\\
				IGR J18293--1213 & 277.33400 & --12.21408      & \nustar & 30161002002 & 2015-09-11 & 24.7 & \citet{clavel2016} \\
						&&& \swift/XRT & 00081763001 && 1.9 & \\
				IGR J20569+4940  & 314.17808 & +49.66850    & \nustar & 30161001002 & 2015-11-13 & 30.3 & This work\\
						&&& \swift/XRT & 00081762001 && 1.9 &\\
				\hline \hline
		\end{tabular}
        \begin{flushleft}
        	$^*$ The X-ray observations of IGR\,J14091--6108 are not listed here because they were not part of this Legacy program but of a joint \xmm, VLT and \nustar\ Guest Observer program \citep[see][]{tomsick2016b}.\\
        	$^{**}$ Coordinates of the soft X-ray counterpart  provided by \chandra\ or \swift/XRT (see Section~\ref{sec:nustardata}).
        \end{flushleft}
\end{table*}

\subsection{\nustar}
\label{sec:nustardata}
We reduced the \nustar\ data using NuSTARDAS v.1.7.1, which is part of HEASOFT v.6.20, and CALDB version 20161021, setting \texttt{saamode=strict} and \texttt{tentacle=yes} in order to better remove the time intervals having an enhanced count rate due to the contamination created by the South Atlantic Anomaly (SAA).

We extracted the source light curves and spectra from a circular region having a 60$''$ radius and centered on the most precise soft X-ray position of each source provided by either \chandra\ for IGR\,J10447--6027 \citep{fiocchi2010} and IGR\,J17402--3656 \citep{tomsick2009b}, or \swift/XRT for IGR\,J16181--5407 \citep{landi2012b} and IGR\,J20569+4940 \citep{landi2010a}, and listed in Table~\ref{tab:obs}. The background light curves and spectra were extracted from a circular region having a 100$''$ radius and located at the other end of the chip in which the source was detected. 

Each \nustar\ spectrum was grouped to reach at least a 5\,$\sigma$ significance in each energy bin, except for the highest energy bins for which we have a lower significance (2.2\,$\sigma$ on average). For illustration purposes, the 5\,$\sigma$ requirement was raised to 10\,$\sigma$ for all \nustar\ spectra presented in Fig.~\ref{fig:IGRJ10447_spec}, \ref{fig:IGRJ16181_spec}, \ref{fig:IGRJ17402_spec}, \ref{fig:IGRJ20569_spec} and \ref{fig:unfoldedspec}.

\subsection{Neil Gehrels \swift\ Observatory}
The \swift/XRT was operated in Photon Counting (PC) mode and the corresponding data were reduced using HEASOFT v.6.20. 
For each observation, we extracted a spectrum from a 30$''$ radius centered on the targeted source.  We also made a background spectrum using a source-free annulus with inner and outer radii of 60$''$--300$''$ (IGR\,J10447--6027), 90$''$--300$''$ (IGR\,J16181--5407) or 120$''$--360$''$ (IGR\,J20569+4940). We used the most recent response matrix for a spectrum in PC mode (swxpc0to12s6\_20130101v014.rmf), and we used {\ttfamily xrtmkarf} with an exposure map to make the ancillary response file.  

Each \swift/XRT spectrum was grouped to obtain at least 25 counts (for IGR\,J20569+4940) or a 2\,$\sigma$ significance (for IGR\,J10447--6027 and IGR\,J16181--5407) in each energy bin. IGR\,J17402--3656 has no simultaneous \swift/XRT coverage due to Moon constraints for the \swift\ spacecraft during the \nustar\ observation.

\subsection{Spectral analysis}
\label{sec:analysis}

\begin{figure}
	\centering
	\includegraphics[width=0.5\textwidth]{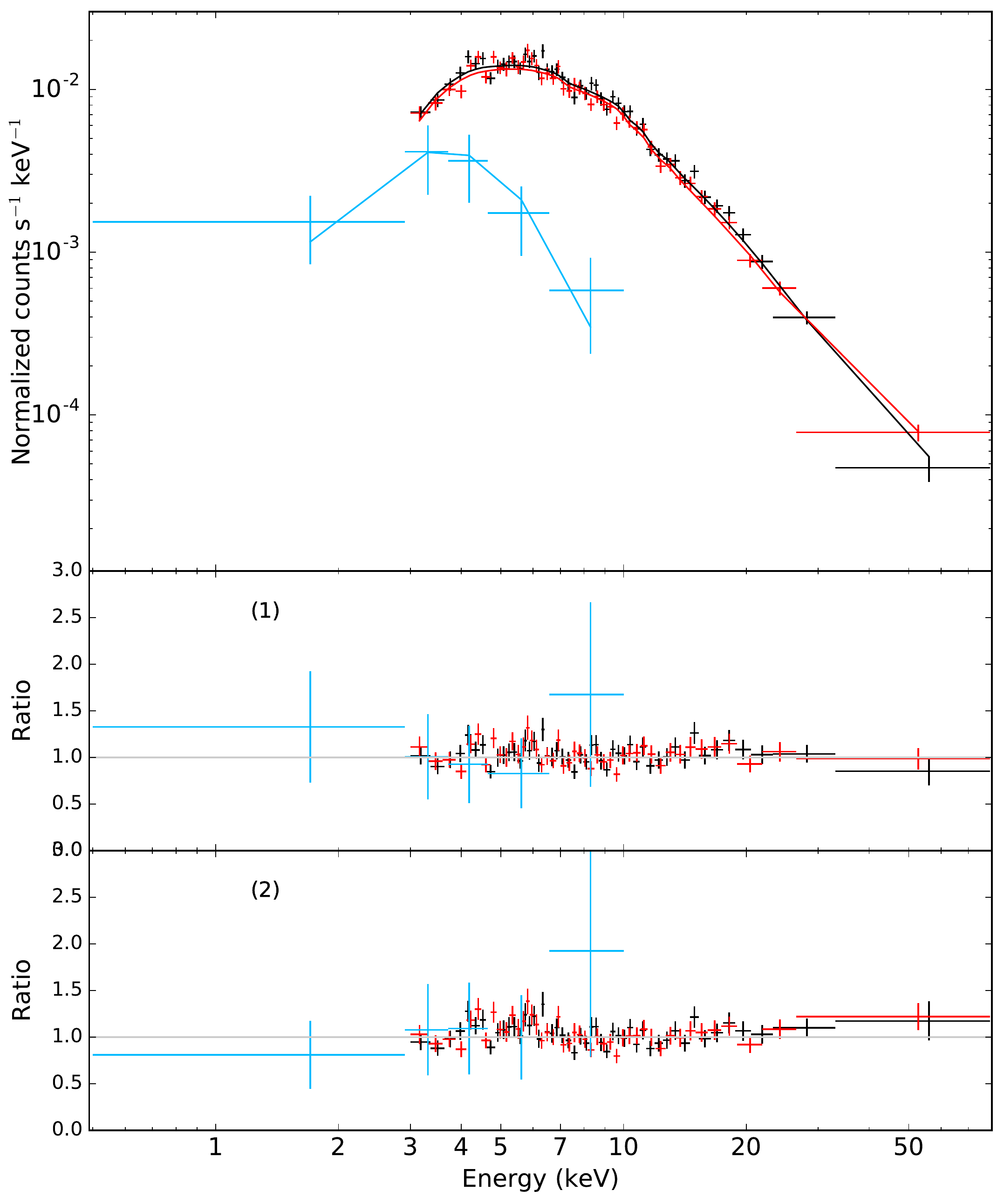}
	\caption{\nustar\ and \swift/XRT spectra of IGR J16181--5407 adjusted by an absorbed power law (top) and residuals associated to Model (\ref{eq:po}, middle) and (\ref{eq:brem}, bottom), for Obs.\ ID 30161006002 (FPMA in black, FPMB in red and XRT in blue). The corresponding best fit parameters are listed in Table~\ref{tab:fitsNS}.}
	\label{fig:IGRJ16181_spec}
\end{figure}

\begin{figure}
	\centering
	\includegraphics[width=0.5\textwidth]{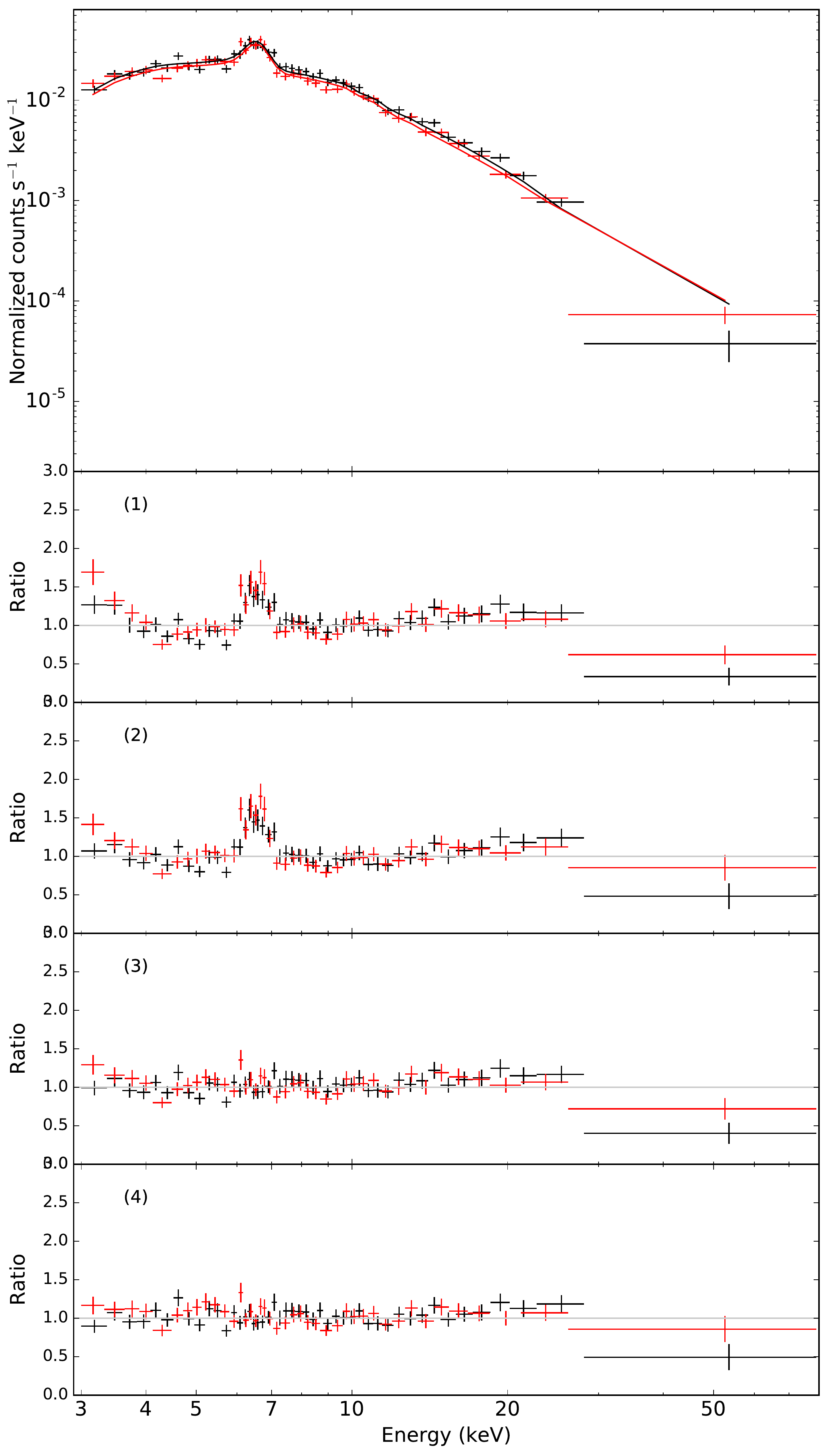}
	\caption{\nustar\ spectra of IGR J17402--3656 adjusted by an absorbed Bremstrahlung plus a Gaussian line (top) and residuals associated to Model (\ref{eq:po}, middle top), (\ref{eq:brem}, middle), (\ref{eq:gauss}, middle bottom) and (\ref{eq:cutoff}, bottom), for Obs.\ ID 30161004002 (FPMA in black and FPMB in red). The corresponding best fit parameters are listed in Table~\ref{tab:fitsNS}.}
	\label{fig:IGRJ17402_spec}
\end{figure}

For each source, all spectra we obtained were fitted simultaneously, using \textsc{xspec} v.12.9.1, with the following two models:
\begin{equation}
	\rm \textsc{const} * \textsc{tbabs} * \textsc{pegpwrlw}
	\label{eq:po}
\end{equation}
\begin{equation}
	\rm \textsc{const} * \textsc{tbabs} * \textsc{bremss}
	\label{eq:brem}
\end{equation}
Additional and more complex models were also tested, including:
 \begin{equation}
	\rm \textsc{const} * \textsc{tbabs} * (\textsc{bremss} + \textsc{gauss})
	\label{eq:gauss}
\end{equation}
\begin{equation}
	\rm \textsc{const} * \textsc{tbabs} * (\textsc{cutoffpl} + \textsc{gauss})
	\label{eq:cutoff}
\end{equation}
Results obtained with the four models listed above are shown in Table~\ref{tab:fitsNS} and in Figures~\ref{fig:IGRJ10447_spec}, \ref{fig:IGRJ16181_spec}, \ref{fig:IGRJ17402_spec} and \ref{fig:IGRJ20569_spec}. All error bars are given with 90\% significance.

We note that the \swift/XRT calibration constant (C$_{Swift}$ in Table~\ref{tab:fitsNS}) is between $\sim$40 and 80\%, i.e.\ lower than expected for the cross-calibration error between \nustar\ and \swift/XRT \citep[see e.g.][]{madsen2017}. This discrepancy can be partly explained by the shortness of the \swift/XRT exposure compared to the \nustar\ one, making it more sensitive to the short timescale variability of the sources, but this is likely not the only reason. However, careful investigation of our data selection and reduction procedure did not provide any obvious cause for the discrepancies\footnote{This careful investigation included: performing simultaneous fits in the 3--10\,keV energy range, quantifying the difference in background subtraction, testing the presence of a cutoff when fixing C$_{Swift}$ to 1, and searching for additional point sources in \nustar\ extraction regions.}. In order to check the impact of the possibly erroneous \swift/XRT data on the fit results, we first fixed C$_{Swift}$ to be equal to 1. In this case, the fits worsen, but the parameters remain consistent with the ones presented in Table~\ref{tab:fitsNS}. Then, we also performed spectral fits of the \nustar\ data alone. The values obtained for Models (\ref{eq:po}) and (\ref{eq:brem}) are once again within the error bars of the ones presented in Table~\ref{tab:fitsNS}, except for IGR\,J20569+4940, which gives a higher column density (\nh\ $= (6.2\pm0.9)\times10^{22}$\,cm$^{-2}$) and a steeper power law ($\Gamma = 2.70\pm0.04$).

Finally, for two sources identified as AGN, we tested the presence of a neutral Compton reflection component by fitting:
\begin{equation}
	\rm \textsc{const} * \textsc{tbabs} * \textsc{pexmon}
	\label{eq:pex}
\end{equation}
and for the source identified as a CV/IP, we adjusted the \nustar\ spectrum with the physical model \textsc{IPM} developed by \cite{suleimanov2005}, also including reflection, partial absorption, and the iron fluorescent lines at 6.4 and 6.7\,keV:
\begin{equation}
	\rm \textsc{const} * \textsc{tbabs} * \textsc{pcfabs} * (\textsc{reflect} * \textsc{ipm} + \textsc{gauss} + \textsc{gauss}).
	\label{eq:ipm}
\end{equation}
This allows the source parameters to be directly compared to the two previous CV/IPs that are among the faint persistent IGR sources and were observed by \nustar\ \citep[see][for more details]{clavel2016,tomsick2016b}.

\begin{figure*}
	\centering
	\includegraphics[width=1.0\textwidth]{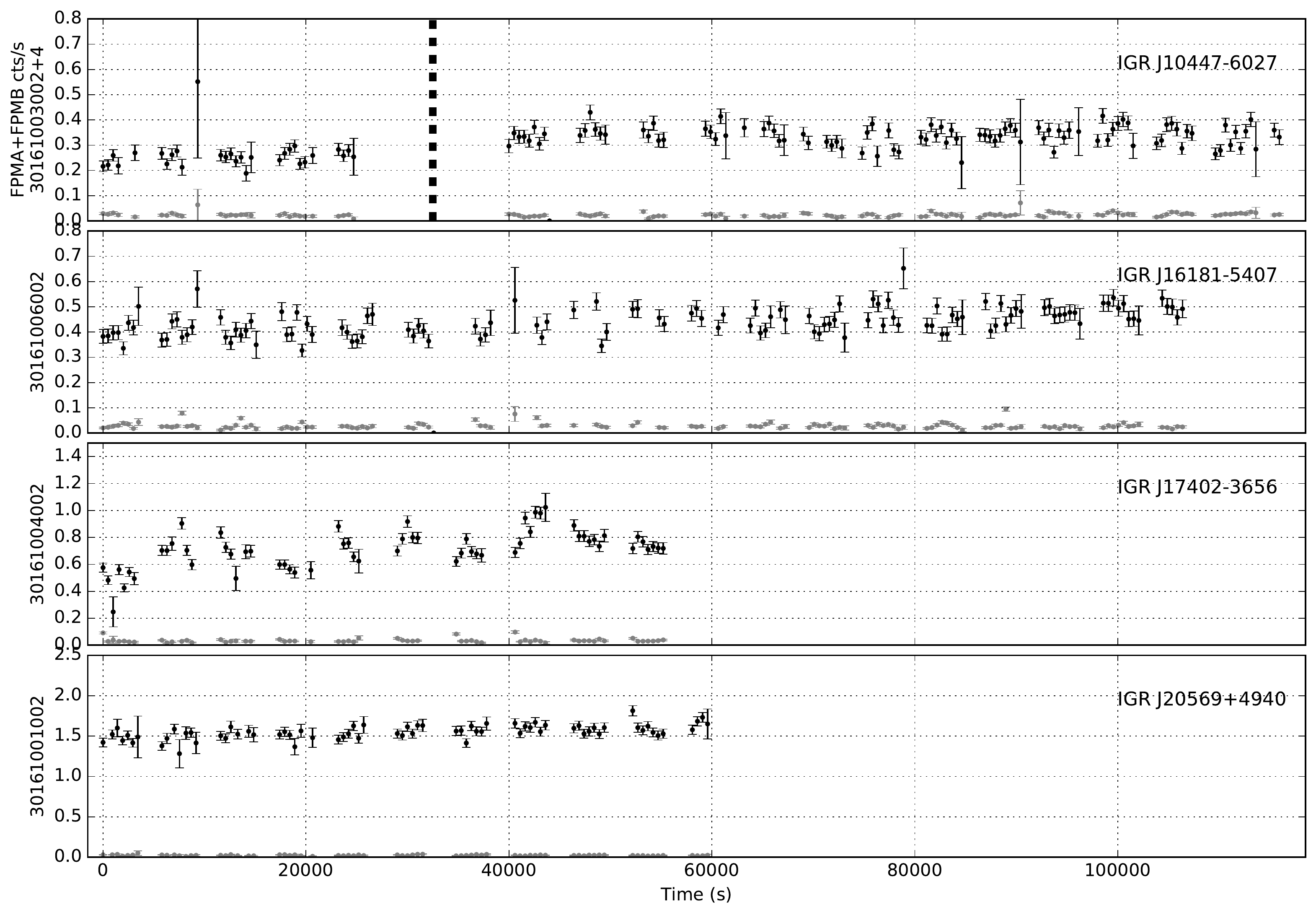}
	\caption{\textit{NuSTAR} lightcurves integrated between 3 and 79\,keV with a 500\,s binning, for the IGR sources (black) and their corresponding background level (gray). Each panel corresponds to one obs.\ ID with the reference time being the beginning of the observation, except for the top panel which includes both observations of IGR\,J10447--6027 (30161003002 starting from 0\,s and 30161003004 arbitrarily starting from 40000\,s) separated by the bold dashed line. Strictly simultaneous \swift/XRT observations (not shown) are only available for obsid 30161003004 (at t\,$\sim$\,56--58\,ks, top panel) and 30161001002 (at t\,$\sim$\,11--13\,ks, bottom panel). The others are either slightly after or before the \nustar\ ones (30161003002 and 30161006002, respectively).}
	\label{fig:lc}

\end{figure*}

\begin{table*} 
        \centering
        \caption{\nustar\ and \swift/XRT simultaneous spectral fits using \textsc{xspec} models presented in Section~\ref{sec:analysis}.}
        \label{tab:fitsNS}
		\begin{tabular}{l l l c c c c }
				\hline \hline
				Model & Param.$^\dagger$ & Unit & IGR J10447--6027$^\star$  & IGR J16181--5407  & IGR J17402--3656 & IGR J20569+4940\\
				\hline
				(\ref{eq:po})	& $\chi^2$/d.o.f. &	& 275.1/271 	& 350.4/332 		& 516.0/302					& 526.6/485 \\
								& \nh 					& $10^{22}$\,cm$^{-2}$	& $16.7\pm2.6$	& $9.7\pm1.7$ 	& $17.8^{+2.4}_{-2.2}$	&  $3.5\pm0.4$\\
								& $\Gamma$ 		&	& $1.50\pm0.06$	& $1.71\pm0.05$ & $1.96\pm 0.06$  		& $2.60\pm0.03$\\
								& F$_{\rm 3-79keV}$					& 10$^{-12}$\,erg\,cm$^{-2}$\,s$^{-1}$ & 	16.9 & 16.4 			& 24.9 						& 33.0\\
								& C$_{\rm Obs2}$		&	& $0.7\pm0.1$ & --- & --- & --- \\
								& C$_{Swift}$		&	&  $0.4\pm0.2$ & $0.6\pm0.2$ & --- & $0.81\pm0.07$\\
				\hline
				(\ref{eq:brem}) & $\chi^2$/d.o.f. & & 267.2/271	& 376.3/332 		& 	476.6/302  		& 675.0/485\\
								& \nh 								& $10^{22}$\,cm$^{-2}$	 & $13.2\pm2.0$	& $4.6\pm1.2$ 	& $9.3\pm1.6$ 	& $1.5\pm0.2$  \\
								& kT 								& keV & $>65.7$	& $40.8^{+6.3}_{-5.1}$ & $25.3^{+2.8}_{-2.4} $ & $7.0\pm0.2$\\
								& F$_{\rm 3-79keV}$								& 10$^{-12}$\,erg\,cm$^{-2}$\,s$^{-1}$ & 	15.1	& 14.0 			& 20.8			& 26.3 \\
								& C$_{\rm Obs2}$				& & $0.7\pm0.1$ & --- & --- & --- \\
								& C$_{Swift}$ 					& & 	$0.4\pm0.2$	& $0.5\pm0.2$ 	& --- 					& $0.86\pm0.08$\\			
				\hline
				(\ref{eq:gauss})	& $\chi^2$/d.o.f. 								&	& & & 	308.5/299  & \\
								& \nh														& $10^{22}$\,cm$^{-2}$		 & &  & $6.0\pm1.6$ & \\
								& kT 														& keV	& &  & $33.2^{+4.5}_{-3.7}$ & \\
								& E$_{\rm line}$ 									& keV	& & & $6.55\pm0.05$  &\\
								& $\sigma_{\rm line}$ 							& keV	& & & $0.27\pm0.06$ &\\
								& N$_{\rm line}$ 								& 10$^{-5}$ ph\,cm$^{-2}$\,s$^{-1}$	& & & $5.9\pm0.8$ & \\
								& F$_{\rm 3-79keV}$ 													& 10$^{-12}$\,erg\,cm$^{-2}$\,s$^{-1}$	& &  & 22.3 & \\
				\hline							
				(\ref{eq:cutoff})	& $\chi^2$/d.o.f. 	&&&      &			283.8/298 		& 475.2/484 \\
								& \nh 					& $10^{22}$\,cm$^{-2}$	&&     &  		$<1.8$		& $2.7\pm0.3$ \\
								& $\Gamma$ 		&	&&& 		$0.78^{+0.14}_{-0.05}$		& $2.22\pm0.09$\\
								& E$_{\rm cut}$ 		& keV	&&& 			$15.5^{+2.8}_{-2.0}$		& $23.9^{+7.9}_{-4.9}$\\
								& E$_{\rm line}$ 									& keV	& & & $6.53\pm0.05$  & ---\\
								& $\sigma_{\rm line}$ 							& keV	& & & $0.30^{+0.06}_{-0.04}$ & ---\\
								& N$_{\rm line}$ 								& 10$^{-5}$ ph\,cm$^{-2}$\,s$^{-1}$	& & & $6.4\pm0.8$ & --- \\
								& F$_{\rm 3-79keV}$ 					& 10$^{-12}$\,erg\,cm$^{-2}$\,s$^{-1}$ 	&&& 				20.6		&29.6\\
								& C$_{Swift}$		&	&&& --- & $0.80\pm0.07$\\								
				\hline \hline
		\end{tabular}
		\begin{flushleft}
		$^\dagger$ F$_{\rm 3-79keV}$ denotes the flux of the source as measured by \nustar\ between 3 and 79\,keV. \\
        $^\star$ IGR\,J10447--6027 analysis combines the two \nustar\ data sets (Obs.\ 1 and Obs.\ 2 referring to Obs.\ ID 3016003004 and 3016003002, respectively) and the two \swift/XRT spectra (the value obtained for the cross-calibration constant C$_{Swift}$ is the same for both observations; this is why a single value is reported in this table). 
        \end{flushleft}
\end{table*}

%%%%%%%%%%%%%%%%%%%%%%%%%%%%%%%%%%%%%%%%%%%%%%%%%%%%%%%%%%%%%%%%%%%%
%%%%%%%%%%%%%%%%%%%%%%%%%%%%%%%%%%%%%%%%%%%%%%%%%%%%%%%%%%%%%%%%%%%%
%%%%%%%%%%%%%%%%%%%%%%%%%%%%%%%%%%%%%%%%%%%%%%%%%%%%%%%%%%%%%%%%%%%%
%%%%%%%%%%%%%%%%%%%%%%%%%%%%%%%%%%%%%%%%%%%%%%%%%%%%%%%%%%%%%%%%%%%%
\section{Four sources observed by \nustar}
\label{sec:nustar}

At the time of the \nustar\ observations, all targets of our Legacy program were unidentified. For each source, we review the information that has now been published and present the \nustar\ results. In addition, information we found by searching the VizieR database\footnote{VizieR service DOI: \href{https://www.doi.org/10.26093/cds/vizier}{10.26093/cds/vizier}} regarding the counterparts of IGR\,J10447--6027 and  IGR\,J17402--3656 is reported in Sections~\ref{sec:J10447sup} and \ref{sec:J17402sup}, and we publish a Keck/LRIS optical spectrum of IGR\,J20569+4940 in Section~\ref{sec:opti}.

%%%%%%%%%%%%%%%%%%%%%%%%%%%%%%%%%%%%%%%%%%%%%%%%%%%%%%%%%%%%%%%%%%%%
%%%%%%%%%%%%%%%%%%%%%%%%%%%%%%%%%%%%%%%%%%%%%%%%%%%%%%%%%%%%%%%%%%%%
\subsection{IGR\,J10447--6027}
\label{sec:J10447}
The soft X-ray and NIR (2MASS\,J10445192--6025115) counterparts of this source have been identified \citep{landi2010a,fiocchi2010}. In particular, previous soft X-ray observations with \swift\ and \chandra\ show a strongly absorbed X-ray source with no sign of variability. The \chandra\ spectrum can be fit with an absorbed power law (\nh\ $=(22\pm3)\times10^{22}$\,cm$^{-2}$, $\Gamma=1.0^{+0.3}_{-0.6}$) for a total unabsorbed flux of F$_{\rm 0.3-10keV} \sim 1.7\times10^{-12}$\,erg\,cm$^{-2}$\,s$^{-1}$. More recently, NIR spectroscopy from VLT/ISAAC provided a redshift measurement of this source \citep[$z=0.047\pm0.001$,][]{fortin2018} demonstrating the extragalactic nature of IGR\,J10447--6027, possibly a Seyfert 2 AGN.  

\subsubsection{NuSTAR results}
\label{sec:src1nu}
IGR\,J10447--6027 was observed twice by \nustar\ as part of the Legacy program: the first observation was cut short (12\,ks) because of a ToO, and it was rescheduled entirely (42\,ks) two months later (see Table~\ref{tab:obs}).

The corresponding light curves are shown in Figure~\ref{fig:lc} (top panel). 
Each \nustar\ observation shows a steady 3--79\,keV emission from the source but there is a $\sim$50\,\% flux increase between Obs.\ 1 (net count rate = 0.09\,cts\,s$^{-1}$\,module$^{-1}$) and Obs.\ 2 (0.14\,cts\,s$^{-1}$\,module$^{-1}$). The fluxes measured by \swift/XRT are consistent with being the same for all observations (two from the \nustar\ Legacy program and the 2007 one by \citealt{landi2010a}). However, note that the first \nustar\ observation does not have simultaneous \swift/XRT coverage (since it started after our \nustar\ observation had been stopped to begin the ToO observation). 

\begin{figure}
	\centering
	\includegraphics[width=0.5\textwidth]{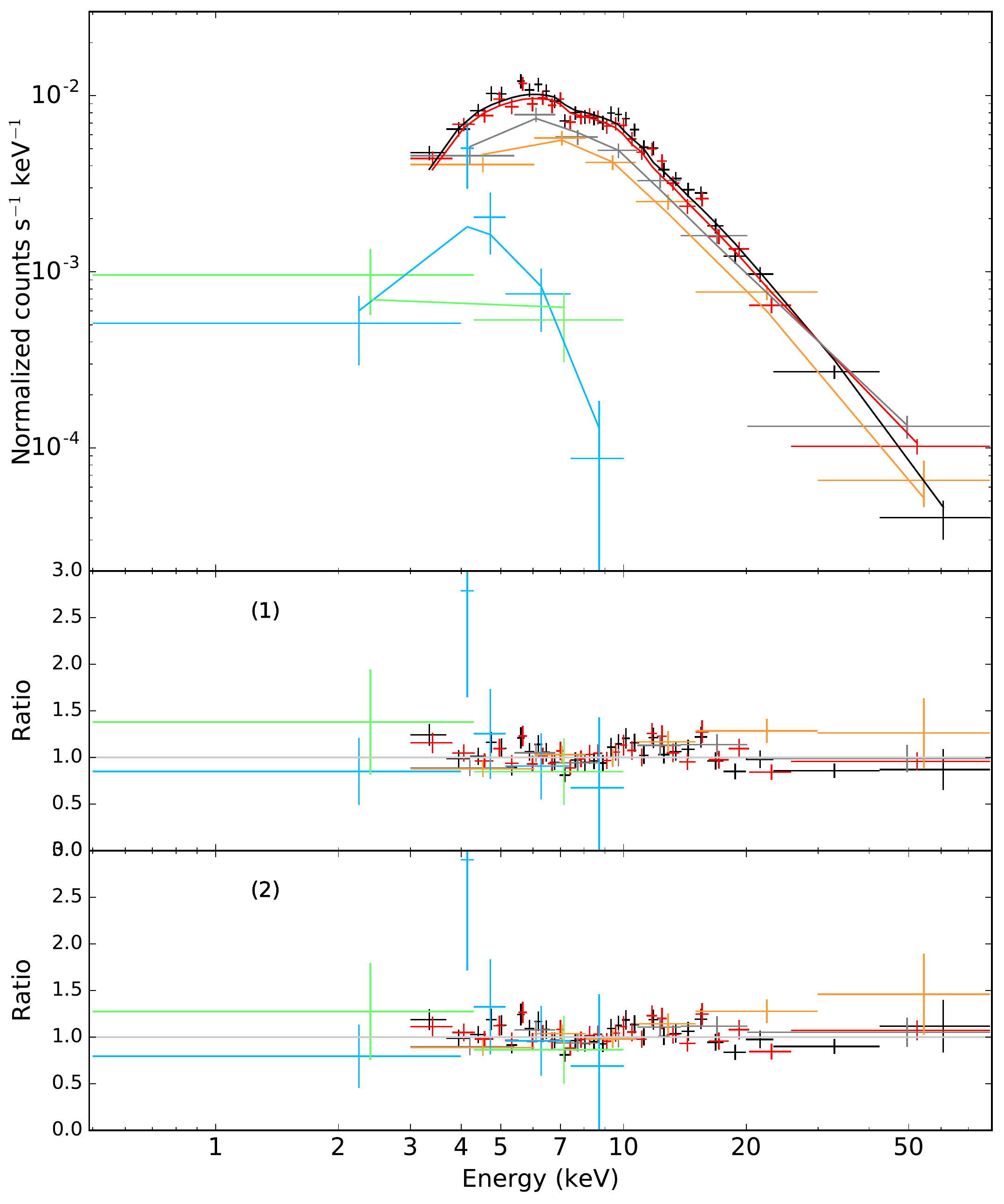}
	\caption{\nustar\ and \swift/XRT spectra of IGR J10447--6027 adjusted by an absorbed power law (top) and residuals associated to Model (\ref{eq:po}, middle) and (\ref{eq:brem}, bottom), for Obs.\ ID 30161003002 (FPMA in gray, FPMB in orange and XRT in green) and Obs.\ ID 30161003004 (FPMA in black, FPMB in red and XRT in blue). The corresponding best fit parameters are listed in Table~\ref{tab:fitsNS}.}
	\label{fig:IGRJ10447_spec}
\end{figure}

The \nustar\ and \swift/XRT spectra obtained in 2015 are all shown in Figure~\ref{fig:IGRJ10447_spec}. When Obs.\ 1 and 2 are fitted separately with an absorbed power law the two data sets give parameters which are consistent within the 90\% error bars (Obs.\ 1: \nh\ $= 24^{+9}_{-8}\times10^{22}$\cm, $\Gamma=1.44\pm0.15$; Obs.\ 2: \nh\ $= (16\pm3)\times10^{22}$\cm, $\Gamma=1.51\pm0.06$). The simultaneous fit of the two observations with all parameters fixed to be the same and the free cross-calibration constant (C$_{Obs2}$) also accounting for the flux difference is satisfactory (see Table~\ref{tab:fitsNS}). The spectra are then best fit by an absorbed power law  
with photon index $\Gamma\sim1.5$ with no sign of emission line features. 
However, the apparent compatibility of the two observations may be due to the poor statistics of Obs.\ 1 preventing an independent fit of both the \nh\ and the power law index (large error bars) and the larger statistics of Obs.\ 2 dominating the simultaneous fit. To test this, we also perform a simultaneous fit of all spectra leaving either \nh\ or $\Gamma$ free to vary between the two observations. The results are statistically equivalent,  ($\chi^2$/d.o.f. = 259.0/270, $\Gamma=1.50\pm0.06$, \nh$_{,\rm Obs1} = 27^{+6}_{-5}\times 10^{22}$\,cm$^{-2}$ and \nh$_{,\rm Obs2} = 15^{+3}_{-2}\times10^{22}$\,cm$^{-2}$) or ($\chi^2$/d.o.f. = 261.3/270,  \nh\ $ = 27^{+6}_{-5}\times 10^{22}$\,cm$^{-2}$, $\Gamma_{\rm Obs1}=1.31\pm0.09$ and $\Gamma_{\rm Obs2}=1.53\pm0.06$), and demonstrate that the spectral shape of IGR\,J10447--6027 is harder in the first observation. 

IGR\,J10447--6027 has a photon index $\Gamma\sim1.5$ consistent with what has been observed for Seyfert galaxies \citep[see e.g.][]{bianchi2009,ricci2017} and a high intrinsic absorption (the line of sight absorption is \nh$_{\rm, Gal}=1.5\times10^{22}$\,cm$^{-2}$ for this source\footnote{Sum of the hydrogen contribution given by the  Leiden/Argentine/Bonn Survey of Galactic H$_{\rm I}$ \citep{kalberla2005} and the molecular contribution derived from the CO map provided by \citet{dame2001}.}), confirming the Seyfert 2 nature of this AGN. In this context, it is likely that the evolution between the two \nustar\ observations (flux increase and spectral softening) are due to a decrease of the column density of the AGN torus along the line of sight \citep[see e.g.][]{marinucci2013}. 

To test the presence of a reflection component we also fit Obs.\ 2 with model (\ref{eq:pex}), fixing all \textsc{pexmon} parameters to their default value, except for $z=0.047$ and $i=60^\circ$. There is no significant improvement of the fit and we obtain an upper limit of $R_{\rm refl} < 0.22$ for the reflection fraction\footnote{The exact value of this upper limit varies with the inclination; it is between $R<0.10$ for $i=0^\circ$ and $R<0.78$ for $i=85^\circ$.}, so the presence of such component is not to be excluded.  

\subsubsection{Supplementary information}
\label{sec:J10447sup}
The position of the 2MASS counterpart of IGR\,J10447--6027 is also consistent with a single \textit{WISE} source \citep[AllWISE J104451.90--602511.5,][]{cutri2014}. This source presents an infrared excess ($W1-W2=1.245$ and $W2-W3=2.407$) which is also typical of AGNs \citep[e.g.][]{stern2012,assef2018,karasev2018}.

%%%%%%%%%%%%%%%%%%%%%%%%%%%%%%%%%%%%%%%%%%%%%%%%%%%%%%%%%%%%%%%%%%%%
%%%%%%%%%%%%%%%%%%%%%%%%%%%%%%%%%%%%%%%%%%%%%%%%%%%%%%%%%%%%%%%%%%%%
\subsection{IGR\,J16181--5407}
\label{sec:J16181}
The \swift/XRT counterpart was identified by \citet{landi2012b}. It is located at ${\rm RA} = 16^h18^m08.02^s$, ${\rm Dec} = -54^\circ06'09.8''$ (6$''$ uncertainty), consistent with three USNO/2MASS sources. The short \swift/XRT exposure did not provide enough statistics to derive spectral information, but its flux was $F_{\rm 2-10keV} = 4\times10^{-13}$\,erg\,cm$^{-2}$\,s$^{-1}$ (assuming a power law with $\Gamma=1.8$ and Galactic absorption \nh$_{,\rm Gal}=0.6\times10^{22}$\,cm$^{-2}$). The recent \chandra\ observation covering this source also pinpoints the position of its soft X-ray counterpart which is consistent with a single \textit{WISE} source having an infrared excess \citep{ursini2018}.

\subsubsection{\nustar\ results}
\label{sec:src2nu}
The \nustar\ lightcurve of IGR\,J16181--5407 is shown in Figure~\ref{fig:lc} (top middle panel), the emission is steady through the observation (less than $\sim20$\% variability in the count rate).

The corresponding spectra are best fit by an absorbed power law with a photon index $\Gamma\sim1.7$ and a high intrinsic absorption \nh$_{, \rm intr}\sim1\times 10^{23}$\,cm$^{-2}$ (see Table~\ref{tab:fitsNS}). This spectral shape is consistent with the source being a Seyfert 2 AGN. In this context, the order of magnitude increase between 2012 \citep{landi2012b} and 2017 ($F_{\rm 2-10keV} = 3.9\times10^{-12}$\,erg\,cm$^{-2}$\,s$^{-1}$) could be due to a variation of \nh\ along the line of sight (from $\sim1.6\times10^{24}$\,cm$^{-2}$ down to $\sim1\times10^{23}$\,cm$^{-2}$). Such variations have been observed in several known AGNs \citep[see e.g.][and references therein]{ricci2016}.

The fit residuals show no obvious sign of emission line features, but we tested the presence of a reflection component by fitting model~(\ref{eq:pex}) on the \nustar\ and \swift/XRT spectra, fixing all parameters to default values except for the inclination angle which is fixed to $i=60^\circ$ and the redshift which is left free to vary. This component slightly improves the fit ($\chi^2$/d.o.f. = 334.9/330, F$_{\rm Test}$ probability less than $6\times10^{-4}$) and the best fit parameters are then \nh\ $=10.2^{+1.7}_{-1.6}\times 10^{22}$\,cm$^{-2}$, $\Gamma=1.86\pm0.08$, $R = 0.57^{+0.32}_{-0.27}$ and $z=0.08^{+0.03}_{-0.02}$.  The quality of the spectrum does not allow us to fit the source inclination. Fixing it at other values in the 0--85$^\circ$ range does change the constraint on the reflection fraction (going as low as $R=0.37^{+0.21}_{-0.17}$ for $i=0^\circ$) but the redshift remains at the same value for all inclinations. This redshift measurement is also in full agreement with the value derived from an independent analysis of the \nustar\ data set, using a Gaussian line to fit the reflection component \citep{ursini2018}. These are further hints that this source is extragalactic. Therefore, based on the \nustar\ observation, IGR\,J16181--5407 is classified as an AGN, in agreement with previously published information.

%%%%%%%%%%%%%%%%%%%%%%%%%%%%%%%%%%%%%%%%%%%%%%%%%%%%%%%%%%%%%%%%%%%%
%%%%%%%%%%%%%%%%%%%%%%%%%%%%%%%%%%%%%%%%%%%%%%%%%%%%%%%%%%%%%%%%%%%%
\subsection{IGR\,J17402--3656 or IGR\,J17404--3655}
\label{sec:J17402}

IGR\,J17404--3655 \citep[R.A. = 265.112$^\circ$, Dec. = --36.913$^\circ$;][]{bird2007} is referenced as an X-ray binary in SIMBAD\footnote{\url{http://simbad.u-strasbg.fr/simbad/}}, and IGR\,J17402--3656 \citep[R.A. = 265.087$^\circ$, Dec. = --36.936$^\circ$;][]{krivonos2007} is referenced as the open cluster NGC6400 in SIMBAD. However, from an \integral\ point of view, these two names refer to a single source, hereafter IGR\,J17402--3656 \citep[e.g.][]{krivonos2012}. Its soft X-ray \citep[CXOU\,J174026.8--365537,][]{tomsick2009b}, NIR \citep[2MASS 17402685--3666374,][]{landi2008} and optical \citep[USNO-A2.0 0525--28851523,][]{masetti2009} counterparts have been identified. 

The optical spectrum only shows a single narrow H$\alpha$ emission line superimposed onto a red continuum, which is typical of X-ray binaries. Furthermore, its optical colors do not match any star of early spectral type (which might rule out an HMXB), and its spectral appearance does not show the numerous emission lines expected in dwarf nova CVs. Based on this information, and the fact that the source is likely distant (substantial spectral absorption), \cite{masetti2009} identified IGR\,J17402--3656 as an LMXB.  
 
The \chandra\ spectrum is very hard \citep[$\Gamma\sim-0.3$,][]{tomsick2009b}, which would be very unusual for an LMXB, and is more commonly seen in either HMXBs that harbor more highly magnetized neutron stars or magnetic CVs. Moreover, the infrared spectrum obtained by \cite{coleiro2013} shows that only the Br$\gamma$ line is detected within the K$_s$-band. This line is mainly detected in supergiant and Be stars, since no He I line was detected at 2.058\,$\mu$m (generally present in supergiant stars), \cite{coleiro2013} identified this source as a Be-HMXB candidate, even if the LMXB scenario could not be ruled out.
 
More recently, deeper infrared spectroscopy did detect the He\,\textsc{i} line, as well as Br$\gamma$ and a weak C\,\textsc{iv} emission line  \citep[VLT/ISAAC,][]{fortin2018}. They argue that the non-detection of Pfund emission rules out a hot B-type star and, based on the NIR study of CVs by \citet{harrison2004}, conclude that IGR\,J17402--3656 could be a CV with a K3--5 V companion if at a distance $d=530$--$700$\,pc.

\subsubsection{\nustar\ results}
The \nustar\ light curve of IGR\,J17402--3656 shows no flaring activity, but there is a factor $\sim2$ variation between minimum and maximum count rates within the observation, consistent with the factor 2 difference observed between the fluxes previously measured by \chandra\ and \swift/XRT \citep[both averaged over short observations,][]{landi2008,tomsick2009b}.

We used the Z$_1^2$ (Rayleigh) test \citep{buccheri1983} to search for signals in the 3--24\,keV lightcurve, making a periodogram extending from 0.0001\,Hz (10\,000\,s) to 10\,Hz. Between 1 and 322.5\,s there are no signals reaching the 3\,$\sigma$ significance threshold (after accounting for trials). Starting from 322.58\,s, there are five peaks above the 3\,$\sigma$ threshold with increasing significance. Monte Carlo simulations of randomly distributed photon arrival times within the \nustar\ GTI, reproduce these signals with the first significant peak also at 322.58\,s. Therefore, these peaks are not due to IGR\,J17402--3656 variability but to the discontinuity of the GTI. Between 0.1 and about 1\,s, the IGR\,J17402--3656 periodogram has 12 peaks slightly above the 3\,$\sigma$ significance threshold (none is above the 5\,$\sigma$ threshold, and only two are above 4\,$\sigma$). 
To further understand the origin of these peaks, we folded the source light curve on 100\,000 frequencies between 0.0001 and 10\,Hz, using 10 phase bins. For each frequency, we then performed a fit of the folded light curve and recorded its amplitude, defined as the difference between maximum and minimum count rates divided by the sum of these two quantities. The folded light curves obtained at the peak frequencies are consistent with being constant ($\chi^2\ll1$) and have amplitudes ranging from 9.0 to 12.2\%.  So the 12 peaks described above are likely signatures of aperiodic variability. Furthermore, 95\% of folded light curves corresponding to periods ranging from 0.1 to 322.58\,s have amplitudes below 9.2\%. Based on this 2\,$\sigma$ upper limit on periodic signals in the 3--24\,keV band, we cannot exclude the presence of a spin pulsation signal such as anticipated for CV/IPs \citep[these spin modulations are energy dependent and their amplitude in the \nustar\ band is limited, see e.g.][]{tomsick2016b}.

The \nustar\ spectra are best fit by an absorbed thermal component, having a temperature $kT \sim 30$\,keV, plus a strong emission line around 6.55\,keV (equivalent width $EW \sim 640$\,eV). The overall absorption fitted with this simple model ($N_{\rm H} \sim 5\times10^{22}$\,cm$^{-2}$) is higher than the value expected for Galactic absorption only (\nh$_{,\rm Gal} = 0.6\times10^{22}$\,cm$^{-2}$)\footnote{Derived using H$_{\rm I}$ and CO surveys (see Section~\ref{sec:src1nu}).}. The temperature found is too high for an LMXB \citep[e.g.][]{tauris2006} while the properties of the emission line do not match what has been observed in HMXBs \citep{torrejo2010,gime2015}. Instead, this spectrum is fully consistent with what is expected for a magnetic CV \citep[e.g.][]{mukai2017}. To further investigate the physical parameters of this Galactic source, we adjust its spectrum with model~(\ref{eq:ipm}). The best fit parameters are listed in Table~\ref{tab:polar}, and are fully consistent with other known IPs, such as IGR\,J18293--1213, which was also identified thanks to the \nustar\ Legacy program \citep{clavel2016}.

\begin{table}
        \centering
        \caption{Spectral parameters obtained by fitting the IPM model to the \nustar\ spectra of IGR\,J17402--3656 ($\chi^2$/d.o.f. = 284.4/297). The uncertainties listed correspond to 90\% confidence intervals. The normalization constant is fixed to 1 for FPMA, and is $1.00\pm0.03$  for FPMB. 
        The X-ray flux of the source is $2.0\times10^{-11}$\,erg\,cm$^{-2}$\,s$^{-1}$ in the 3--79\,keV range.}
        \label{tab:polar}
        \renewcommand{\arraystretch}{1.5}
        \begin{tabular}{l c c c}
        \hline \hline
        Model & Param.$^*$ & Unit & Best Fit  \\
        \hline
		\textsc{tbabs} & \nh & $10^{22}$\,cm$^{-2}$ &  $0.6$ (fixed) \\
		\textsc{pcfabs} & \nh$_{,\rm pc}$ &  $10^{22}$\,cm$^{-2}$ &  $45^{+18}_{-16}$ \\ 
					&  fraction & --- &  $0.52\pm0.06$\\ \hline
		\textsc{reflect}  & $\Omega/2\pi$ & --- & $1.0$ (fixed)\\
					& A & --- & $>0.55$\\ 
					&A$_{\rm Fe}$& --- & $1.0$ (fixed)\\
					& cos\,$\alpha$ & --- & $>0.42$\\ \hline 
		\textsc{ipm} & $M_{\rm wd}$ &\msol & $0.73^{+0.08}_{-0.07}$\\ 
					 & F$_{3-79\rm keV}$& $10^{-3}$\phcms & $1.3\pm0.3$\\\hline 
		\textsc{gauss} & E$_{\rm Fe\,K\alpha}$ & keV & $6.4$ (fixed)\\
					   & $\sigma_{\rm Fe\,K\alpha}$ & eV & $50$ (fixed)\\
					   & N$_{\rm Fe\,K\alpha}$ & $10^{-5}$\phcms & $2.6\pm0.8$\\ \hline 
		\textsc{gauss} & E$_{\rm Fe\,xxv}$ & keV & $6.7$ (fixed)\\
					   & $\sigma_{\rm Fe\,xxv}$ & eV & $50$ (fixed)\\ 
					   & N$_{\rm Fe\,xxv}$ & $10^{-5}$\phcms & $ 2.6\pm0.7$\\ 
        \hline \hline
        \end{tabular}
        \begin{flushleft}
        $^*$ Parameters of the \textsc{reflect} model are: the reflection scaling factor ($\Omega/2\pi$, set to 1 for an isotropic source above a disk), the abundance of elements heavier than He relative to solar (A), the iron abundance relative to the previous one (A$_{\rm Fe}$) and the inclination angle of the white dwarf magnetic field ($\alpha$). In the \textsc{ipm} model, M$_{\rm wd}$ refers to the mass of the white dwarf.
		\end{flushleft}
\end{table}

\subsubsection{Supplementary information}   
\label{sec:J17402sup}
The \chandra\ counterpart of IGR\,J17402--3656 also corresponds to a single \textit{Gaia} source having a parallax measurement of $0.38\pm0.14$\,mas \citep[Gaia DR2\footnote{The other four targets of this Legacy program (Table~\ref{tab:obs}) do not have reliable parallax measurements due to either an absence of optical counterparts or the presence of a significant astrometric excess noise.},][]{gaiacollaboration2018}. This translates into a distance $d=2.6^{+1.9}_{-0.8}$\,kpc \citep{bailerjones2018}, significantly greater than the one anticipated by \citet{fortin2018}. At 2.6\,kpc, IGR\,J17402--3656 would have an X-ray luminosity $L_{\rm 3-79keV}\sim1.6\times10^{34}$\,erg\,s$^{-1}$, which is too high for a Polar \citep[e.g.][]{sazonov2006} and fully compatible with the average luminosity derived from IPs with known distances \citep[e.g.][]{schwope2018, suleimanov2019}. In this scenario, the IR luminosity of the IP system is likely to be dominated by the emission of the accretion disk, explaining why it could be about one order of magnitude higher than what has been anticipated for the companion star by \citet{fortin2018}.

%%%%%%%%%%%%%%%%%%%%%%%%%%%%%%%%%%%%%%%%%%%%%%%%%%%%%%%%%%%%%%%%%%%%
%%%%%%%%%%%%%%%%%%%%%%%%%%%%%%%%%%%%%%%%%%%%%%%%%%%%%%%%%%%%%%%%%%%%
\subsection{IGR\,J20569+4940}
\label{sec:J20569}
\citet{landi2010a} identified the soft X-ray \citep[later named 1SXPS\,J205642.6+494009 in \swift\ catalogs, e.g.][]{evans2014}, NIR (2MASS\,J20564271+4940068) and radio (NVSS\,205642+494005) counterparts of IGR\,J20569+4940. In the radio, the source is bright and has a 2.8 to 11\,cm flat spectral index of --0.4 \citep{reich2000}, making it a likely radio-loud object, i.e. either a microquasar or a blazar \citep{landi2010a}. This source is also associated with a \fermi\ source \citep[2FGL\,J2056.7+4939,][]{marti2012} which was classified as a blazar candidate in the last \fermi\ catalog \citep[3LAC\,J2056.7+4938,][]{ackermann2015} and it has recently been detected in the TeV energy range by VERITAS \citep{mukherjee2016}.

IGR\,J20569+4940 also appears to be variable both in radio \citep[listed in the Variable 1.4GHz radio sources from NVSS and FIRST,][]{ofek2011} and in X-rays \citep[fluxes different by a factor 2 reported for XMMSL1\,J205642.7+494004 on the timescale of hours,][]{landi2010a}.

\subsubsection{NuSTAR results}

The \nustar\ light curve of IGR\,J20569+4940, displayed in Figure~\ref{fig:lc} (bottom panel), shows no sign of strong variability.

\begin{figure}
	\centering
	\includegraphics[width=0.5\textwidth]{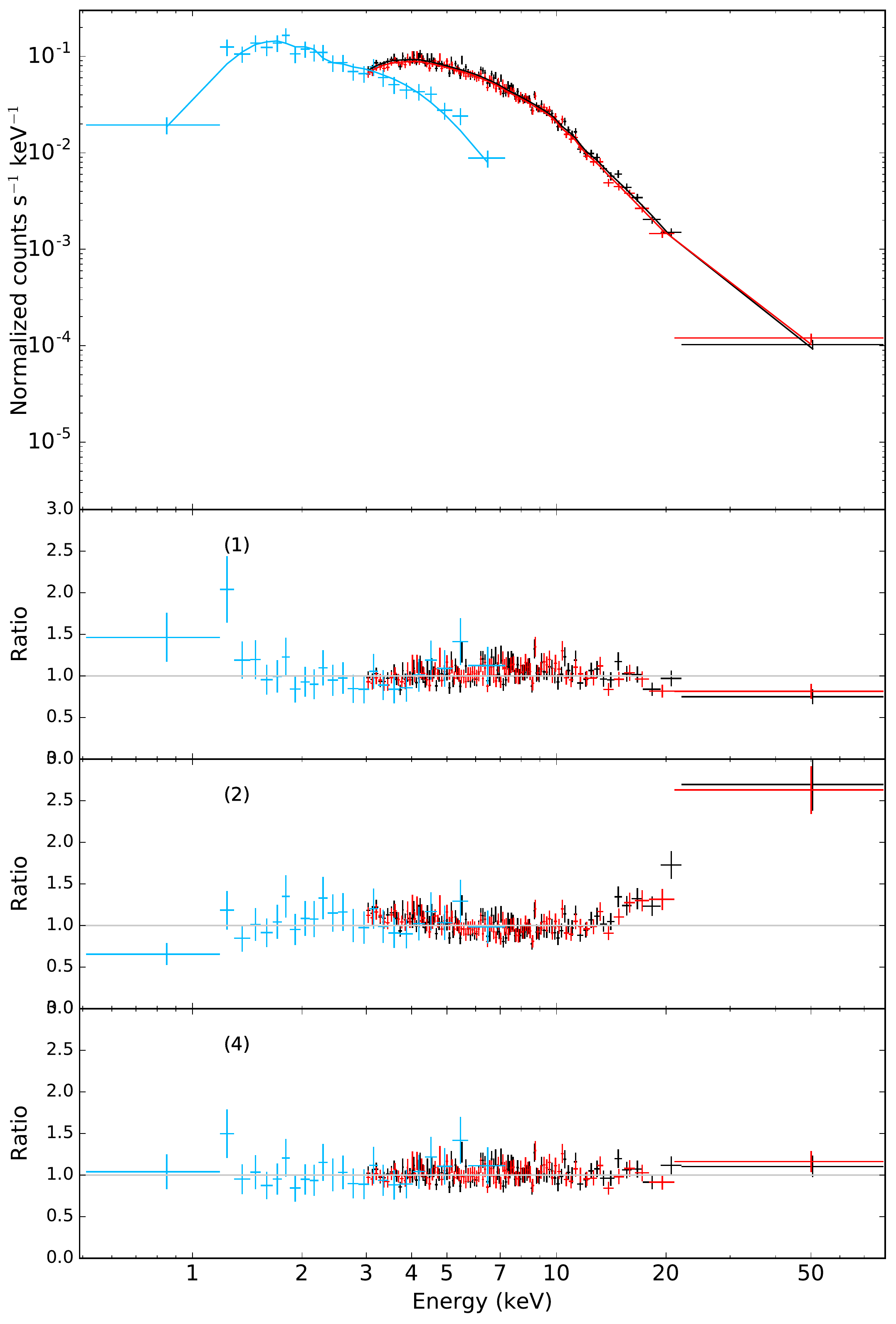}
	\caption{\nustar\ and \swift/XRT spectra of IGR J20569+4940 adjusted by an absorbed cutoff power law (top) and residuals associated to Model (\ref{eq:po}, middle top), (\ref{eq:brem}, middle bottom) and (\ref{eq:cutoff}, bottom), for Obs.\ ID 30161001002 (FPMA in black, FPMB in red and XRT in blue). The corresponding best fit parameters are listed in Table~\ref{tab:fitsNS}.}
	\label{fig:IGRJ20569_spec}
\end{figure}

The corresponding \nustar\ and \swift/XRT spectra are shown in Figure~\ref{fig:IGRJ20569_spec}. They are best fit by an absorbed cutoff power law (see Table~\ref{tab:fitsNS}), revealing a rather steep hard X-ray spectrum. The \nustar\ data set is indeed compatible with an absorbed power law of photon index $\Gamma\sim2.7$ ($\chi^2$/d.o.f. = 479.8/462). Such a spectrum and the absence of emission feature is an additional sign that we are seeing the synchrotron emission from the jet of a blazar up to the hard X-ray energy range. In this case, we would expect the presence of a second emission bump, created by Comptonisation, in the $\gamma$-ray energy range. We do not detect the spectral turnover in our data set. However, the \textit{Fermi} spectrum \citep[F$_{\rm1-100keV}=1.30\pm0.19\times10^{-9}$\,ph\,cm$^{-2}$\,s$^{-1}$ with $\Gamma=1.78\pm0.08$,][]{acero2015} indicates that the spectrum has indeed two bumps. This is consistent with the source being a blazar and the absence of optical emission line (see Sect.~\ref{sec:opti}) makes IGR\,J20569+4940 a BL Lac \citep[see e.g.][]{ghisellini2017}.

The power law index we measured by the cutoff power law is compatible with the one previously measured by \swift/XRT in 2009 \citep{landi2010a}. However, the total column density is higher in 2016 (\nh\ $=2.7\pm0.3\times10^{22}$\,cm$^{-2}$) than it was in 2009 (\nh\ $=1.53^{+0.18}_{-0.16}\times10^{22}$\,cm$^{-2}$), and the 2--10\,keV flux is also about twice higher in 2016 than in 2009. Known BL Lacs have a wide variety of variability patterns with timescales ranging from intraday to years and amplitudes going from few percents to orders of magnitude \citep[e.g.][]{rani2017,kapanadze2018b,kapanadze2018a}. So IGR\,J20569+4940 moderate variability on long timescale and the absence of strong variations in the \nustar\ light curve are both consistent with its classification as a BL Lac. 

\subsubsection{Optical spectrum}
\label{sec:opti}
\begin{figure}
	\centering
	\includegraphics[width=0.5\textwidth]{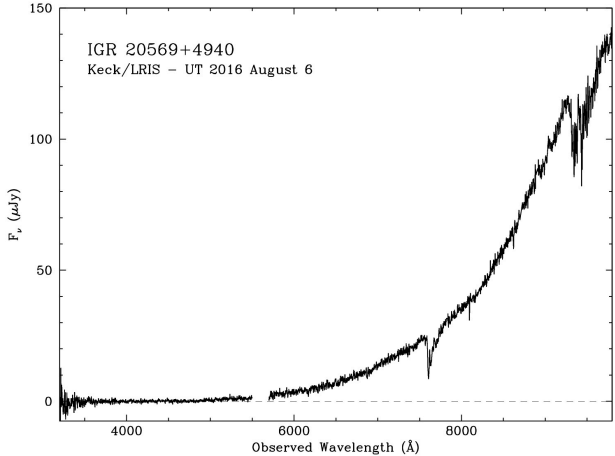}
	\caption{Keck/LRIS optical spectrum of IGR J20569+4940.}
	\label{fig:IGRJ20569_keck}
\end{figure}

In complement of the \nustar\ Legacy program we obtained a spectrum from Keck/LRIS from two 900\,s integration on UT 2016 August 6 (PI: F.~Harrison) centered on the candidate optical counterpart to IGR\,20569+4940 (see Figure~\ref{fig:IGRJ20569_keck}). 
The object is extremely red, and featureless (other than telluric absorption). 
This is compatible with a highly absorbed BL Lac optical spectrum, as expected from both the high Galactic absorption in the direction of IGR\,J20569+4940  (\nh$_{,\rm Gal}=1\times10^{22}$\,cm$^{-2}$) and the column density derived from the X-ray spectra. 

\section{Discussion}
\label{sec:synthesis}

The four classifications described in Section~\ref{sec:nustar} illustrate the multi-wavelength effort dedicated to the follow up of faint hard X-ray sources detected by \integral. The \nustar\ Legacy program `unidentified \integral\ sources' was designed to be part of this effort, and to test the identification diagnostics provided by the \nustar\ observations. With the aim of better understanding the population of faint HMXBs, our initial target selection was the sample of twenty-six sources listed as unidentified by \citet[][see Table~\ref{tab:cat}]{lutovinov2013}. As of today, six of these sources have been observed by \nustar\, five through our Legacy program (see Table~\ref{tab:obs}) and one through an independent program \citep{tomsick2016b}. Using the results provided by these six observations, we first discuss how \nustar\ can be efficiently used to investigate the population of faint hard X-ray sources in Section~\ref{sec:nuclass}. Then, we combine all classification results now available regarding our initial target selection (Table~\ref{tab:cat}) to give an updated description of the HMXB population at low fluxes (Sections~\ref{sec:hmxb1} and \ref{sec:hmxb2}). For this purpose, any new classification, whether it is an HMXB or not, is improving constraints on the surface density of HMXBs. In this respect, \nustar\ has provided decisive information for about 10\% of the new identifications.

\subsection{Source classification using NuSTAR}
\label{sec:nuclass}
Taking full advantage of the better sensitivity and the higher spectral resolution of \nustar, we have been successfully investigating the fainter population associated with the unidentified \integral\ sources, focusing only on persistent sources \citep[see also][]{lutovinov2013}.

\begin{figure*}
	\centering
	\includegraphics[width=\textwidth]{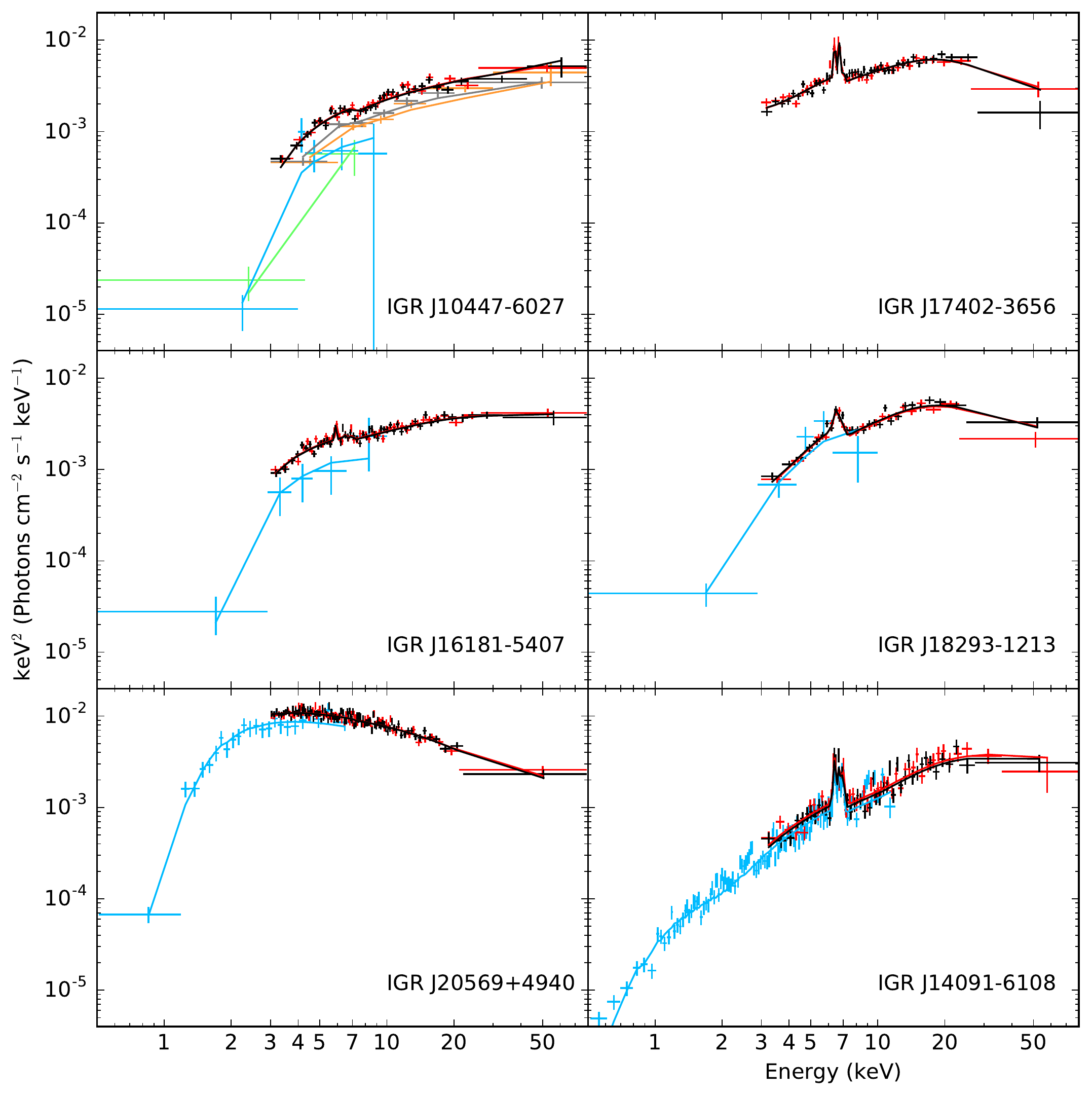}
	\caption{Unfolded spectra obtained for all six sources observed by \nustar: the five observations obtained through the `Unidentified \integral\ sources' Legacy program (including IGR\,J18293--1213, originally published in \citealt{clavel2016}) and one independent observation (IGR\,J14091--6108, originally published in \citealt{tomsick2016b}). The spectral model corresponds to the best fit model: power law (IGR\,J10447--6027), \textsc{pexmon} (IGR\,J16181--5407) and cutoff power law (IGR\,J20569+4940) models for the extragalactic sources (left panels) and IPM model for the Intermediate Polars (right panels). The data correspond to \nustar/FPMA (black and grey), FPMB (red and orange) and \swift/XRT spectra (blue and green), except for IGR\,J14091--6108 where \xmm/pn data is shown (blue).}
	\label{fig:unfoldedspec}
\end{figure*}

\nustar\ provided crucial inputs for the spectral studies of three magnetic CVs \citep[see also][]{tomsick2016b,clavel2016} and three AGNs. The corresponding unfolded spectra are presented in Figure~\ref{fig:unfoldedspec}. The spectral coverage extending up to $\sim80$\,keV allowed for a constraint on the intrinsic hardness of the sources with, for instance, measurements of Bremsstrahlung temperatures above $kT\gtrsim15$\,keV for all sources classified as IPs (Fig.~\ref{fig:unfoldedspec}, right panels). The iron line properties also revealed to be a compelling \nustar\ diagnostic for our study, since they are particularly prominent in magnetic CVs (Fig.~\ref{fig:unfoldedspec}, right panels) and a reshifted reflected emission was also detected in one of the extragalactic sources (see Sect.~\ref{sec:src2nu}). 
In addition, variability studies performed on the \nustar\ light curves were powerful tools to describe the properties of faint IGR sources, especially the detection of eclipses in IGR\,J18293--1213 \citep{clavel2016} and the confirmation of both a pulsation period and a decrease of the pulse fraction with energy in IGR\,J14091--6108 \citep{tomsick2016b}. Therefore, combined with observations at other wavelengths, \nustar\ can be efficiently used to classify faint hard X-ray sources towards the Galactic plane and to better constrain their individual characteristics. 

Among the sample observed by \nustar, three CVs were identified, which is not surprising since this population of Galactic sources is expected to be dominant at low fluxes \citep[see e.g.][]{lutovinov2013}. The individual properties of these three sources are consistent with the overall population of hard X-ray sources already classified as CV/IPs \citep[see e.g.][]{pretorius2014,schwope2018}, and constraints on their distance demonstrate that they are relatively fainter only because they are further away. 
Moreover, the classification of IGR\,J14091--6108, IGR\,J17402--3656 and IGR\,J18293--1213 as CV/IPs  highlights the importance of X-ray diagnostics to pinpoint the right identification for this type of source, whereas optical or infrared spectral information alone may be misleading (see e.g.\ Sect.~\ref{sec:J17402}). Therefore, deep X-ray observations would help classify the remaining unidentified IGR sources listed Table~\ref{tab:cat}, including IGR\,J18256--1035, which current classification as an LMXB candidate \citep[][mostly based on its optical spectrum]{masetti2013a} is in contradiction with the hard spectrum measured by \chandra\ \citep{tomsick2008}. These diagnostics are similar to the ones we originally had for IGR\,J17402--3656, which has now been classified as a CV/IP based on its \nustar\ spectrum.

\subsection{Constraining the population of faint HMXBs}
\label{sec:hmxb1}

\begin{figure*}
	\centering
	\includegraphics[width=\textwidth]{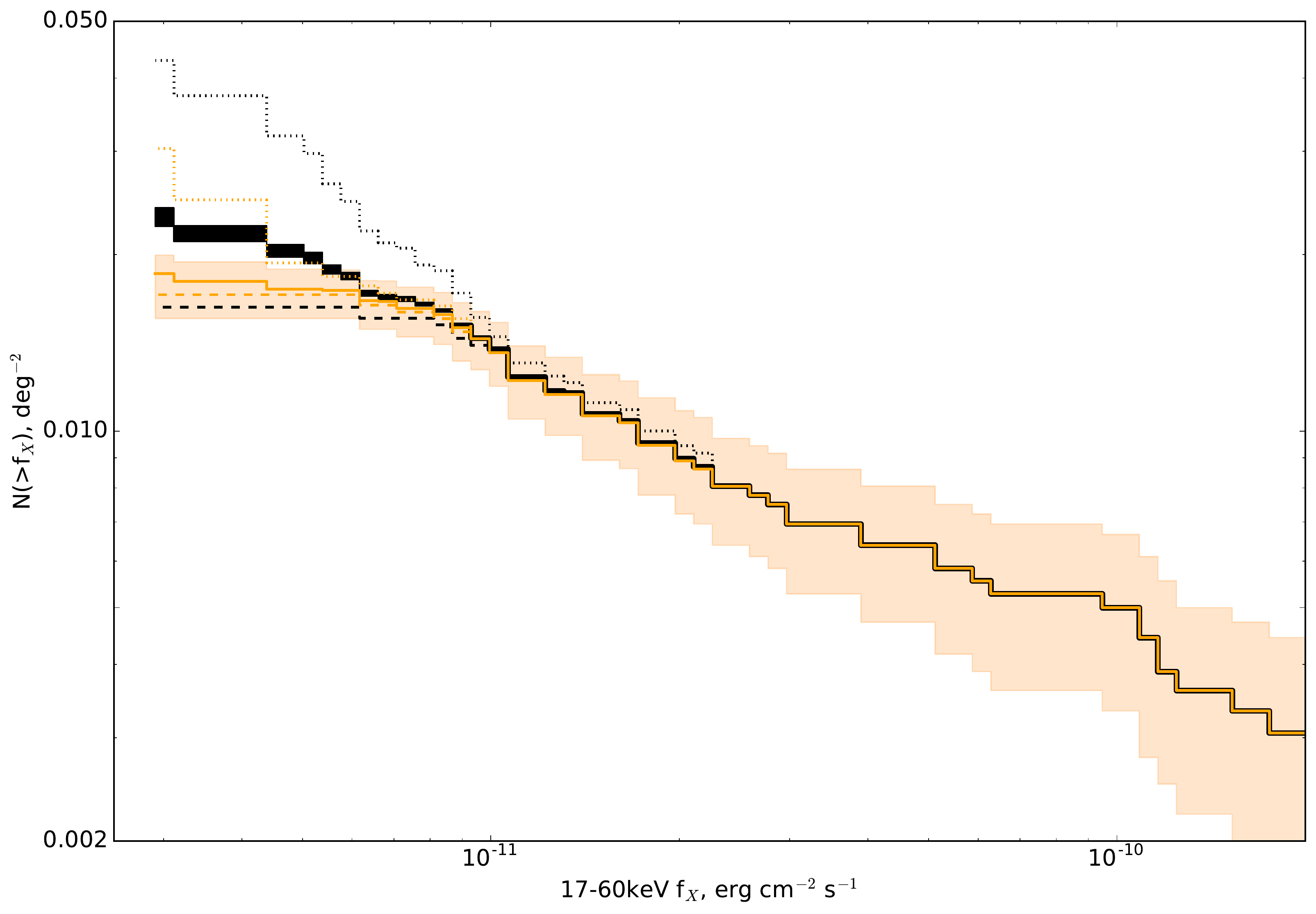}
	\caption{Surface density of HMXBs in the Galactic plane \citep[sky coverage provided by the 9-year \integral\ survey and restricted to $\vert b \vert <5^\circ$,][]{krivonos2012}. \textit{(Black)} Results from \citet{lutovinov2013} when 26 persistent sources were unidentified. \textit{(Orange)} Results from this work with 6 persistent sources still unidentified. Solid lines show the best predictions from \citet[][at least 12 HMXBs are among all IGR persistent sources with $F_{\rm 17-60keV} < 10^{-11}$\,erg\,cm$^{-2}$\,s$^{-1}$, bold black line]{lutovinov2013} and from this work (unidentified sources are HMXBs with a probability of 50\% for XTE\,J1824--141 and 10\% for all others, orange line).
	Extreme cases, with either no HMXB among unidentified sources (dashed lines) or all unidentified sources being HMXBs (dotted lines) are also shown for comparison. IGR\,J17586--2129, a former HMXB candidate now identified as a likely CV candidate \citep{fortin2018}, is not included in any of these curves but the dotted black one. We point out that the sky coverage $A(f_X)$ is not defined for $f_X<3.2\times 10^{-12}$\,erg\,cm$^{-2}$\,s$^{-1}$, so the normalization of the lowest flux bin is arbitrary.
 The orange shading is the bootstrap 95\% confidence interval, the upper limit was computed using the current prediction (solid orange line) and the lower limit using secured identifications (dashed orange line). } 
	\label{fig:lognlogs}
\end{figure*}

The successful identification of several faint persistent IGR sources summarized in Table~\ref{tab:cat} allows for new constraints on the surface density of HMXBs towards low fluxes. To derive the log$N$-log$S$ from the properties of known persistent HMXBs, we use the following equation, which also accounts for the incompleteness of the source classification: 
\begin{equation}
N\left(>f_X\right) = \sum_{\substack{i=1\\f_i > f_X}}^{n_{\rm src}} \frac{P_i}{A\left(f_i\right)},
\label{eq:N}
\end{equation}
where $n_{src}$ is the total number of IGR sources considered as HMXBs, HMXB candidates or unidentified, $P_i$ is the probability of each source $i$ of flux $f_i$ to be an HMXB and $A$ is the sky coverage corresponding to the \integral\ catalog used for the present study \citep[][but restricted to $\vert b \vert < 5^\circ$]{krivonos2012}. In this work, we consider all 53 HMXBs listed in \citet[][Table~1]{lutovinov2013} to have $P_i=1$. The probability $P_i$ of the former HMXB candidate IGR\,J17586--2129\footnote{IGR\,J17586--2129 is listed as a candidate HMXB in \citet{krivonos2012}. It has a flux above $F_{\rm 17-60keV} = 10^{-11}$\,erg\,cm$^{-2}$\,s$^{-1}$ but because its distance was unknown, it was not used by \citet{lutovinov2013} to describe the luminosity function of persistent HMXBs in our Galaxy. So, this new identification should not impact their results and predictions.}, recently identified as a CV \citep{fortin2018}, is instead set to zero. Then we add the twenty previously unidentified sources listed in Table~\ref{tab:cat} with the following probabilities: $P_i=1$ for the two sources now identified as HMXBs, $P_i=0$ for the eighteen sources for which the persistent HMXB identification is ruled out, $P_i = 0.5$ for the source identified as a pulsar (either high or low mass X-ray binary) and, following our HMXB detection rate, $P_i=0.1$ for the remaining five unidentified sources (see Table~\ref{tab:cat}). The result is shown as an orange line in Figure~\ref{fig:lognlogs}. The 95\% confidence interval corresponding to our best prediction was computed using the bootstrap method. To show the full range allowed for $N(>f_X)$ by the present data set, extreme cases where the pulsar and the five unidentified sources are all set to either HMXBs ($P_i = 1$) or non-HMXBs ($P_i=0$) are also displayed in Figure~\ref{fig:lognlogs}. 

We compare our results with the continuous extension of the luminosity function of persistent HMXBs anticipated by \citet{lutovinov2013} in the flux range that can be investigated with \integral. Based on this hypothesis, they predicted that there should be at least 12 HMXBs having a flux below $F_{\rm 17-60keV} < 1\times 10^{-11}$\,erg\,cm$^{-2}$\,s$^{-1}$ within their sample, and possibly up to 14\% more, when correcting their function for the number of known HMXBs with unknown distances that were not included in their study. This gives a probability between $P_i=6/26$ and $P_i=7.68/26$ for the twenty-six sources that were unidentified at the time. The corresponding range of surface densities is reported by the bold black line in Figure~\ref{fig:lognlogs}, significantly above our best prediction. 

\subsection{Is there a deficit of HMXBs at low fluxes?}
\label{sec:hmxb2}
Six sources, including the pulsar XTE\,J1824--141, are still unidentified. We cannot exclude that four of these sources could be HMXBs, in which case their surface density would be in full agreement with \citet{lutovinov2013}'s prediction. However, given the ratio of HMXBs currently identified among the faint IGR sources, this scenario seems unlikely. Instead, the confidence interval associated with what we believe to be a reasonable prediction (one out of the six unidentified sources is expected to be an HMXB) is showing a significant deviation from the continuous extension proposed by \citet{lutovinov2013}, pointing towards a lower number of persistent HMXBs towards low fluxes. If this is confirmed, the break in the log$N$-log$S$ could be right below $F_{\rm 17-60keV} \sim 1\times 10^{-11}$\,erg\,cm$^{-2}$\,s$^{-1}$ (see Figure~\ref{fig:lognlogs}).

Multi-wavelength efforts to identify the nature of IGR\,J17315--3221, AX\,J1753.5--2745, XTE\,J1824--141, IGR\,J18497--0248,  IGR\,J19113+1413 and Swift\,J2037.2+4151, will be key to test the trend described above. However, to fully understand accretion processes occurring in persistent HMXBs, it will be important to also enlarge the sample of classified faint hard X-ray sources. In this respect, the latest \integral\ Galactic plane survey catalog contains a set of 72 newly detected sources, including 59 below $F_{\rm 17-60keV} < 1\times10^{-11}$\,\ergcms\ \citep{krivonos2017}. Most of these faint sources have no clear identification yet, but once they do, adding this sample to our study will also improve the description of the log$N$-log$S$ for $F_{\rm 17-60keV} =  10^{-12}$--$10^{-11}$\,\ergcms.

In parallel, \nustar\ has also been used to constrain the surface density of hard X-ray sources below the detection limit of \integral. The \nustar\ serendipitous survey and the survey of the Norma arm region led to the detection of four new HMXB candidates. Nevertheless, with no secure classification, it is also not yet possible to discriminate between the different predictions regarding the number of persistent HMXBs at these even fainter fluxes \citep{fornasini2017,tomsick2017}.

\section{Conclusion}
\label{sec:conclu}
The \nustar\ Legacy program titled `Unidentified \integral\ sources' investigates the population of faint and persistent hard X-ray sources revealed by \integral\ in the Galactic plane. As of 2012--2013, twenty-six faint and persistent sources were unidentified \citep{krivonos2012,lutovinov2013}. Out of these twenty-six sources, twenty have now been identified, six using dedicated \nustar\ observations and fourteen by independent efforts (see Table~\ref{tab:cat} and Figure~\ref{fig:distrib}). About half of the persistent sources identified are extragalactic, and about half of the Galactic ones are Intermediate Polars. So far, only two high mass X-ray binaries have been identified in this sample, raising the number of known HMXBs among the persistent IGR sources with fluxes $F_{\rm 17-60keV} < 1\times10^{-11}$\ergcms\ to eight.

The source classification is still incomplete, but based on the twenty sources now identified, we can make reasonable assumptions about the number of HMXBs we expect among the remaining unidentified sources, allowing us to set new constraints on the surface density of faint persistent HMXBs in the Galactic plane. Our work shows a tentative deviation from what has been predicted in previous studies \citep{lutovinov2013}, with possibly fewer HMXBs below $F_{\rm 17-60keV} =10^{-11}$\ergcms. Nevertheless, this trend strongly depends on the hypothesis made to account for the remaining unidentified sources and could become negligible if several HMXBs are found among them. Therefore, it is essential to investigate this further by classifying as many hard X-ray sources as possible both near and below the detection limit of the current \integral\ Galactic plane survey.

\section*{Acknowledgments}
This work was supported under NASA Contract No. NNG08FD60C, and made use of data from the \nustar\ mission, a project led by  the California Institute of Technology, managed by the Jet Propulsion  Laboratory, and funded by the National Aeronautics and Space Administration. We thank the \nustar\ Operations, Software and  Calibration teams for support with the execution and analysis of these observations.  The optical data presented were obtained at the W.\ M.\ Keck Observatory, which is operated as a scientific partnership among the California Institute of Technology, the University of California and the National Aeronautics and Space Administration. The Observatory was made possible by the generous financial support of the W.\ M.\ Keck Foundation.  MC acknowledges financial support from the French National Research Agency in the framework of the ``Investissements d'avenir'' program (ANR-15-IDEX-02) and from CNES, and thanks C.~Babusiaux, G.~Henri and P.-O.~Petrucci for useful discussions. RK acknowledges support from the Russian Science Foundation (grant 19-12-00396). This research has made use of the \nustar\ Data Analysis Software (NuSTARDAS) jointly developed by the ASI Science Data Center (ASDC, Italy) and the California Institute of Technology (USA), of data from the European Space Agency (ESA) mission \textit{Gaia}, processed by the Gaia Data Processing and Analysis Consortium (DPAC), and of the SIMBAD database and the VizieR catalog access tool, CDS, Strasbourg, France.

\software{NuSTARDAS (v.1.7.1), HEAsoft (v6.20; HEASARC 2014), XSPEC (v12.9.1; Arnaud 1996)}




\bibliographystyle{aasjournal}
\bibliography{Clavel_IGR_sources_accepted}

\begin{thebibliography}{}
\expandafter\ifx\csname natexlab\endcsname\relax\def\natexlab#1{#1}\fi
\providecommand{\url}[1]{\href{#1}{#1}}
\providecommand{\dodoi}[1]{doi:~\href{http://doi.org/#1}{\nolinkurl{#1}}}
\providecommand{\doeprint}[1]{\href{http://ascl.net/#1}{\nolinkurl{http://ascl.net/#1}}}
\providecommand{\doarXiv}[1]{\href{https://arxiv.org/abs/#1}{\nolinkurl{https://arxiv.org/abs/#1}}}

\bibitem[{{Abbott} {et~al.}(2018){Abbott}, {Abbott}, {Abbott}, {Abraham},
  {Acernese}, {Ackley}, {Adams}, {Adhikari}, \& et~al.}]{abbott2018}
{Abbott}, B.~P., {Abbott}, R., {Abbott}, T.~D., {et~al.} 2018, arXiv e-prints,
  arXiv:1811.12907.
\newblock \doarXiv{1811.12907}

\bibitem[{{Acero} {et~al.}(2015){Acero}, {Ackermann}, {Ajello}, {Albert},
  {Atwood}, {Axelsson}, {Baldini}, {Ballet}, \& et~al.}]{acero2015}
{Acero}, F., {Ackermann}, M., {Ajello}, M., {et~al.} 2015, \apjs, 218, 23,
  \dodoi{10.1088/0067-0049/218/2/23}

\bibitem[{{Ackermann} {et~al.}(2015){Ackermann}, {Ajello}, {Atwood}, {Baldini},
  {Ballet}, {Barbiellini}, {Bastieri}, {Becerra Gonzalez}, \&
  et~al.}]{ackermann2015}
{Ackermann}, M., {Ajello}, M., {Atwood}, W.~B., {et~al.} 2015, \apj, 810, 14,
  \dodoi{10.1088/0004-637X/810/1/14}

\bibitem[{{Assef} {et~al.}(2018){Assef}, {Stern}, {Noirot}, {Jun}, {Cutri}, \&
  {Eisenhardt}}]{assef2018}
{Assef}, R.~J., {Stern}, D., {Noirot}, G., {et~al.} 2018, \apjs, 234, 23,
  \dodoi{10.3847/1538-4365/aaa00a}

\bibitem[{{Bailer-Jones} {et~al.}(2018){Bailer-Jones}, {Rybizki}, {Fouesneau},
  {Mantelet}, \& {Andrae}}]{bailerjones2018}
{Bailer-Jones}, C.~A.~L., {Rybizki}, J., {Fouesneau}, M., {Mantelet}, G., \&
  {Andrae}, R. 2018, \aj, 156, 58, \dodoi{10.3847/1538-3881/aacb21}

\bibitem[{{Bernardini} {et~al.}(2018){Bernardini}, {de Martino}, {Mukai}, \&
  {Falanga}}]{bernardini2018}
{Bernardini}, F., {de Martino}, D., {Mukai}, K., \& {Falanga}, M. 2018, \mnras,
  478, 1185, \dodoi{10.1093/mnras/sty1090}

\bibitem[{{Bianchi} {et~al.}(2009){Bianchi}, {Guainazzi}, {Matt}, {Fonseca
  Bonilla}, \& {Ponti}}]{bianchi2009}
{Bianchi}, S., {Guainazzi}, M., {Matt}, G., {Fonseca Bonilla}, N., \& {Ponti},
  G. 2009, \aap, 495, 421, \dodoi{10.1051/0004-6361:200810620}

\bibitem[{{Bird} {et~al.}(2007){Bird}, {Malizia}, {Bazzano}, {Barlow},
  {Bassani}, {Hill}, {B{\'e}langer}, {Capitanio}, \& et~al.}]{bird2007}
{Bird}, A.~J., {Malizia}, A., {Bazzano}, A., {et~al.} 2007, \apjs, 170, 175,
  \dodoi{10.1086/513148}

\bibitem[{{Bird} {et~al.}(2016){Bird}, {Bazzano}, {Malizia}, {Fiocchi},
  {Sguera}, {Bassani}, {Hill}, {Ubertini}, \& et~al.}]{bird2016}
{Bird}, A.~J., {Bazzano}, A., {Malizia}, A., {et~al.} 2016, \apjs, 223, 15,
  \dodoi{10.3847/0067-0049/223/1/15}

\bibitem[{{Bogdanov} {et~al.}(2014){Bogdanov}, {Esposito}, {Crawford},
  {Possenti}, {McLaughlin}, \& {Freire}}]{bogdanov2014}
{Bogdanov}, S., {Esposito}, P., {Crawford}, Fronefield, I., {et~al.} 2014,
  \apj, 781, 6, \dodoi{10.1088/0004-637X/781/1/6}

\bibitem[{{Brorby} {et~al.}(2016){Brorby}, {Kaaret}, {Prestwich}, \&
  {Mirabel}}]{brorby2016}
{Brorby}, M., {Kaaret}, P., {Prestwich}, A., \& {Mirabel}, I.~F. 2016, \mnras,
  457, 4081, \dodoi{10.1093/mnras/stw284}

\bibitem[{{Buccheri} {et~al.}(1983){Buccheri}, {Bennett}, {Bignami}, {Bloemen},
  {Boriakoff}, {Caraveo}, {Hermsen}, {Kanbach}, \& et~al.}]{buccheri1983}
{Buccheri}, R., {Bennett}, K., {Bignami}, G.~F., {et~al.} 1983, \aap, 128, 245

\bibitem[{{Clavel} {et~al.}(2016){Clavel}, {Tomsick}, {Bodaghee}, {Chiu},
  {Fornasini}, {Hong}, {Krivonos}, {Ponti}, \& et~al.}]{clavel2016}
{Clavel}, M., {Tomsick}, J.~A., {Bodaghee}, A., {et~al.} 2016, \mnras, 461,
  304, \dodoi{10.1093/mnras/stw1330}

\bibitem[{{Coleiro} {et~al.}(2013){Coleiro}, {Chaty}, {Zurita Heras}, {Rahoui},
  \& {Tomsick}}]{coleiro2013}
{Coleiro}, A., {Chaty}, S., {Zurita Heras}, J.~A., {Rahoui}, F., \& {Tomsick},
  J.~A. 2013, \aap, 560, A108, \dodoi{10.1051/0004-6361/201322382}

\bibitem[{{Cutri} {et~al.}(2014){Cutri}, {Wright}, {Conrow}, {Fowler}, \&
  et~al.}]{cutri2014}
{Cutri}, R.~M., {Wright}, E.~L., {Conrow}, T., {Fowler}, J.~W., \& et~al. 2014,
  VizieR Online Data Catalog, II/328

\bibitem[{{D'A{\`\i}} {et~al.}(2011){D'A{\`\i}}, {La Parola}, {Cusumano},
  {Segreto}, {Romano}, {Vercellone}, \& {Robba}}]{dai2011}
{D'A{\`\i}}, A., {La Parola}, V., {Cusumano}, G., {et~al.} 2011, \aap, 529,
  A30, \dodoi{10.1051/0004-6361/201016401}

\bibitem[{{Dame} {et~al.}(2001){Dame}, {Hartmann}, \& {Thaddeus}}]{dame2001}
{Dame}, T.~M., {Hartmann}, D., \& {Thaddeus}, P. 2001, \apj, 547, 792,
  \dodoi{10.1086/318388}

\bibitem[{{Evans} {et~al.}(2014){Evans}, {Osborne}, {Beardmore}, {Page},
  {Willingale}, {Mountford}, {Pagani}, {Burrows}, \& et~al.}]{evans2014}
{Evans}, P.~A., {Osborne}, J.~P., {Beardmore}, A.~P., {et~al.} 2014, \apjs,
  210, 8, \dodoi{10.1088/0067-0049/210/1/8}

\bibitem[{{Fiocchi} {et~al.}(2010){Fiocchi}, {Bassani}, {Bazzano}, {Ubertini},
  {Landi}, {Capitanio}, \& {Bird}}]{fiocchi2010}
{Fiocchi}, M., {Bassani}, L., {Bazzano}, A., {et~al.} 2010, \apj, 720, 987,
  \dodoi{10.1088/0004-637X/720/2/987}

\bibitem[{{Fornasini} {et~al.}(2017){Fornasini}, {Tomsick}, {Hong}, {Gotthelf},
  {Bauer}, {Rahoui}, {Stern}, {Bodaghee}, \& et~al.}]{fornasini2017}
{Fornasini}, F.~M., {Tomsick}, J.~A., {Hong}, J., {et~al.} 2017, \apjs, 229,
  33, \dodoi{10.3847/1538-4365/aa61fc}

\bibitem[{{Fortin} {et~al.}(2018){Fortin}, {Chaty}, {Coleiro}, {Tomsick}, \&
  {Nitschelm}}]{fortin2018}
{Fortin}, F., {Chaty}, S., {Coleiro}, A., {Tomsick}, J.~A., \& {Nitschelm},
  C.~H.~R. 2018, \aap, 618, A150, \dodoi{10.1051/0004-6361/201731265}

\bibitem[{{Gaia Collaboration} {et~al.}(2018){Gaia Collaboration}, {Brown},
  {Vallenari}, {Prusti}, {de Bruijne}, {Babusiaux}, {Bailer-Jones}, {Biermann},
  \& et~al.}]{gaiacollaboration2018}
{Gaia Collaboration}, {Brown}, A.~G.~A., {Vallenari}, A., {et~al.} 2018, \aap,
  616, A1, \dodoi{10.1051/0004-6361/201833051}

\bibitem[{{Ghisellini} {et~al.}(2017){Ghisellini}, {Righi}, {Costamante}, \&
  {Tavecchio}}]{ghisellini2017}
{Ghisellini}, G., {Righi}, C., {Costamante}, L., \& {Tavecchio}, F. 2017,
  \mnras, 469, 255, \dodoi{10.1093/mnras/stx806}

\bibitem[{{Gim{\'e}nez-Garc{\'\i}a} {et~al.}(2015){Gim{\'e}nez-Garc{\'\i}a},
  {Torrej{\'o}n}, {Eikmann}, {Mart{\'\i}nez-N{\'u}{\~n}ez}, {Oskinova},
  {Rodes-Roca}, \& {Bernab{\'e}u}}]{gime2015}
{Gim{\'e}nez-Garc{\'\i}a}, A., {Torrej{\'o}n}, J.~M., {Eikmann}, W., {et~al.}
  2015, \aap, 576, A108, \dodoi{10.1051/0004-6361/201425004}

\bibitem[{{Hare} {et~al.}(2019){Hare}, {Halpern}, {Clavel}, {Grindlay},
  {Rahoui}, \& {Tomsick}}]{hare2019}
{Hare}, J., {Halpern}, J.~P., {Clavel}, M., {et~al.} 2019, \apj, 878, 15,
  \dodoi{10.3847/1538-4357/ab1cbe}

\bibitem[{{Harrison} {et~al.}(2013){Harrison}, {Craig}, {Christensen},
  {Hailey}, {Zhang}, {Boggs}, {Stern}, {Cook}, \& et~al.}]{harrison2013}
{Harrison}, F.~A., {Craig}, W.~W., {Christensen}, F.~E., {et~al.} 2013, \apj,
  770, 103, \dodoi{10.1088/0004-637X/770/2/103}

\bibitem[{{Harrison} {et~al.}(2004){Harrison}, {Osborne}, \&
  {Howell}}]{harrison2004}
{Harrison}, T.~E., {Osborne}, H.~L., \& {Howell}, S.~B. 2004, \aj, 127, 3493,
  \dodoi{10.1086/420706}

\bibitem[{{Kalberla} {et~al.}(2005){Kalberla}, {Burton}, {Hartmann}, {Arnal},
  {Bajaja}, {Morras}, \& {P{\"o}ppel}}]{kalberla2005}
{Kalberla}, P.~M.~W., {Burton}, W.~B., {Hartmann}, D., {et~al.} 2005, \aap,
  440, 775, \dodoi{10.1051/0004-6361:20041864}

\bibitem[{{Kapanadze} {et~al.}(2018{\natexlab{a}}){Kapanadze}, {Vercellone},
  {Romano}, {Hughes}, {Aller}, {Aller}, {Kharshiladze}, \&
  {Tabagari}}]{kapanadze2018b}
{Kapanadze}, B., {Vercellone}, S., {Romano}, P., {et~al.} 2018{\natexlab{a}},
  \apj, 858, 68, \dodoi{10.3847/1538-4357/aabbac}

\bibitem[{{Kapanadze} {et~al.}(2018{\natexlab{b}}){Kapanadze}, {Vercellone},
  {Romano}, {Hughes}, {Aller}, {Aller}, {Kharshiladze}, {Kapanadze}, \&
  et~al.}]{kapanadze2018a}
---. 2018{\natexlab{b}}, \apj, 854, 66, \dodoi{10.3847/1538-4357/aaa75d}

\bibitem[{{Karasev} {et~al.}(2018){Karasev}, {Lutovinov}, {Tkachenko},
  {Khorunzhev}, {Krivonos}, {Medvedev}, {Pavlinsky}, {Burenin}, \&
  et~al.}]{karasev2018}
{Karasev}, D.~I., {Lutovinov}, A.~A., {Tkachenko}, A.~Y., {et~al.} 2018,
  Astronomy Letters, 44, 522, \dodoi{10.1134/S1063773718090037}

\bibitem[{{Krivonos} {et~al.}(2007){Krivonos}, {Revnivtsev}, {Churazov},
  {Sazonov}, {Grebenev}, \& {Sunyaev}}]{krivonos2007}
{Krivonos}, R., {Revnivtsev}, M., {Churazov}, E., {et~al.} 2007, \aap, 463,
  957, \dodoi{10.1051/0004-6361:20065626}

\bibitem[{{Krivonos} {et~al.}(2012){Krivonos}, {Tsygankov}, {Lutovinov},
  {Revnivtsev}, {Churazov}, \& {Sunyaev}}]{krivonos2012}
{Krivonos}, R., {Tsygankov}, S., {Lutovinov}, A., {et~al.} 2012, \aap, 545,
  A27, \dodoi{10.1051/0004-6361/201219617}

\bibitem[{{Krivonos} {et~al.}(2017){Krivonos}, {Tsygankov}, {Mereminskiy},
  {Lutovinov}, {Sazonov}, \& {Sunyaev}}]{krivonos2017}
{Krivonos}, R.~A., {Tsygankov}, S.~S., {Mereminskiy}, I.~A., {et~al.} 2017,
  \mnras, 470, 512, \dodoi{10.1093/mnras/stx1276}

\bibitem[{{La Parola} {et~al.}(2013){La Parola}, {Cusumano}, {Segreto},
  {D'A{\`\i}}, {Masetti}, \& {D'Elia}}]{laparola2013}
{La Parola}, V., {Cusumano}, G., {Segreto}, A., {et~al.} 2013, \apjl, 775, L24,
  \dodoi{10.1088/2041-8205/775/1/L24}

\bibitem[{{Landi} {et~al.}(2010){Landi}, {Bassani}, {Malizia}, {Stephen},
  {Bazzano}, {Fiocchi}, \& {Bird}}]{landi2010a}
{Landi}, R., {Bassani}, L., {Malizia}, A., {et~al.} 2010, \mnras, 403, 945,
  \dodoi{10.1111/j.1365-2966.2010.16183.x}

\bibitem[{{Landi} {et~al.}(2012){Landi}, {Bassani}, {Masetti}, {Bazzano},
  {Tarana}, \& {Bird}}]{landi2012b}
{Landi}, R., {Bassani}, L., {Masetti}, N., {et~al.} 2012, The Astronomer's
  Telegram, 4233, 1

\bibitem[{{Landi} {et~al.}(2008){Landi}, {Masetti}, {Malizia}, {Del Santo},
  {Tarana}, {Bird}, {Dean}, {Caraveo}, \& et~al.}]{landi2008}
{Landi}, R., {Masetti}, N., {Malizia}, A., {et~al.} 2008, The Astronomer's
  Telegram, 1539, 1

\bibitem[{{Lutovinov} {et~al.}(2013){Lutovinov}, {Revnivtsev}, {Tsygankov}, \&
  {Krivonos}}]{lutovinov2013}
{Lutovinov}, A.~A., {Revnivtsev}, M.~G., {Tsygankov}, S.~S., \& {Krivonos},
  R.~A. 2013, \mnras, 431, 327, \dodoi{10.1093/mnras/stt168}

\bibitem[{{Madsen} {et~al.}(2017){Madsen}, {Beardmore}, {Forster}, {Guainazzi},
  {Marshall}, {Miller}, {Page}, \& {Stuhlinger}}]{madsen2017}
{Madsen}, K.~K., {Beardmore}, A.~P., {Forster}, K., {et~al.} 2017, \aj, 153, 2,
  \dodoi{10.3847/1538-3881/153/1/2}

\bibitem[{{Marinucci} {et~al.}(2013){Marinucci}, {Risaliti}, {Wang}, {Bianchi},
  {Elvis}, {Matt}, {Nardini}, \& {Braito}}]{marinucci2013}
{Marinucci}, A., {Risaliti}, G., {Wang}, J., {et~al.} 2013, \mnras, 429, 2581,
  \dodoi{10.1093/mnras/sts534}

\bibitem[{{Markwardt}(2008)}]{Markwardt2008}
{Markwardt}, C.~B. 2008, The Astronomer's Telegram, 1686, 1

\bibitem[{{Mart{\'\i}} {et~al.}(2012){Mart{\'\i}}, {Luque-Escamilla},
  {S{\'a}nchez-Ayaso}, \& {Mu{\~n}oz-Arjonilla}}]{marti2012}
{Mart{\'\i}}, J., {Luque-Escamilla}, P.~L., {S{\'a}nchez-Ayaso}, E., \&
  {Mu{\~n}oz-Arjonilla}, A.~J. 2012, \apss, 340, 143,
  \dodoi{10.1007/s10509-012-1030-9}

\bibitem[{{Masetti} {et~al.}(2009){Masetti}, {Parisi}, {Palazzi},
  {Jim{\'e}nez-Bail{\'o}n}, {Morelli}, {Chavushyan}, {Mason}, {McBride}, \&
  et~al.}]{masetti2009}
{Masetti}, N., {Parisi}, P., {Palazzi}, E., {et~al.} 2009, \aap, 495, 121,
  \dodoi{10.1051/0004-6361:200811322}

\bibitem[{{Masetti} {et~al.}(2013){Masetti}, {Parisi}, {Palazzi},
  {Jim{\'e}nez-Bail{\'o}n}, {Chavushyan}, {McBride}, {Rojas}, {Steward}, \&
  et~al.}]{masetti2013a}
---. 2013, \aap, 556, A120, \dodoi{10.1051/0004-6361/201322026}

\bibitem[{{Mereminskiy} {et~al.}(2016){Mereminskiy}, {Krivonos}, {Lutovinov},
  {Sazonov}, {Revnivtsev}, \& {Sunyaev}}]{mereminskiy2016}
{Mereminskiy}, I.~A., {Krivonos}, R.~A., {Lutovinov}, A. e.~A., {et~al.} 2016,
  \mnras, 459, 140, \dodoi{10.1093/mnras/stw613}

\bibitem[{{Mukai}(2017)}]{mukai2017}
{Mukai}, K. 2017, \pasp, 129, 062001, \dodoi{10.1088/1538-3873/aa6736}

\bibitem[{{Mukherjee} \& {VERITAS Collaboration}(2016)}]{mukherjee2016}
{Mukherjee}, R., \& {VERITAS Collaboration}. 2016, The Astronomer's Telegram,
  9721, 1

\bibitem[{{Ofek} \& {Frail}(2011)}]{ofek2011}
{Ofek}, E.~O., \& {Frail}, D.~A. 2011, \apj, 737, 45,
  \dodoi{10.1088/0004-637X/737/1/45}

\bibitem[{{Postnov} {et~al.}(2017){Postnov}, {Oskinova}, \&
  {Torrej{\'o}n}}]{postnov2017}
{Postnov}, K., {Oskinova}, L., \& {Torrej{\'o}n}, J.~M. 2017, \mnras, 465,
  L119, \dodoi{10.1093/mnrasl/slw223}

\bibitem[{{Pretorius} \& {Mukai}(2014)}]{pretorius2014}
{Pretorius}, M.~L., \& {Mukai}, K. 2014, \mnras, 442, 2580,
  \dodoi{10.1093/mnras/stu990}

\bibitem[{{Rahoui} {et~al.}(2017){Rahoui}, {Tomsick}, \&
  {Krivonos}}]{rahoui2017}
{Rahoui}, F., {Tomsick}, J.~A., \& {Krivonos}, R. 2017, \mnras, 465, 1563,
  \dodoi{10.1093/mnras/stw2830}

\bibitem[{{Rani} {et~al.}(2017){Rani}, {Stalin}, \& {Rakshit}}]{rani2017}
{Rani}, P., {Stalin}, C.~S., \& {Rakshit}, S. 2017, \mnras, 466, 3309,
  \dodoi{10.1093/mnras/stw3228}

\bibitem[{{Reich} {et~al.}(2000){Reich}, {F{\"u}rst}, {Reich}, {Kothes},
  {Brinkmann}, \& {Siebert}}]{reich2000}
{Reich}, W., {F{\"u}rst}, E., {Reich}, P., {et~al.} 2000, \aap, 363, 141

\bibitem[{{Ricci} {et~al.}(2016){Ricci}, {Bauer}, {Arevalo}, {Boggs}, {Brand
  t}, {Christensen}, {Craig}, {Gandhi}, \& et~al.}]{ricci2016}
{Ricci}, C., {Bauer}, F.~E., {Arevalo}, P., {et~al.} 2016, \apj, 820, 5,
  \dodoi{10.3847/0004-637X/820/1/5}

\bibitem[{{Ricci} {et~al.}(2017){Ricci}, {Trakhtenbrot}, {Koss}, {Ueda}, {Del
  Vecchio}, {Treister}, {Schawinski}, {Paltani}, \& et~al.}]{ricci2017}
{Ricci}, C., {Trakhtenbrot}, B., {Koss}, M.~J., {et~al.} 2017, \apjs, 233, 17,
  \dodoi{10.3847/1538-4365/aa96ad}

\bibitem[{{Sazonov} {et~al.}(2006){Sazonov}, {Revnivtsev}, {Gilfanov},
  {Churazov}, \& {Sunyaev}}]{sazonov2006}
{Sazonov}, S., {Revnivtsev}, M., {Gilfanov}, M., {Churazov}, E., \& {Sunyaev},
  R. 2006, \aap, 450, 117, \dodoi{10.1051/0004-6361:20054297}

\bibitem[{{Schwope}(2018)}]{schwope2018}
{Schwope}, A.~D. 2018, \aap, 619, A62, \dodoi{10.1051/0004-6361/201833723}

\bibitem[{{Stern} {et~al.}(2012){Stern}, {Assef}, {Benford}, {Blain}, {Cutri},
  {Dey}, {Eisenhardt}, {Griffith}, \& et~al.}]{stern2012}
{Stern}, D., {Assef}, R.~J., {Benford}, D.~J., {et~al.} 2012, \apj, 753, 30,
  \dodoi{10.1088/0004-637X/753/1/30}

\bibitem[{{Suleimanov} {et~al.}(2005){Suleimanov}, {Revnivtsev}, \&
  {Ritter}}]{suleimanov2005}
{Suleimanov}, V., {Revnivtsev}, M., \& {Ritter}, H. 2005, \aap, 435, 191,
  \dodoi{10.1051/0004-6361:20041283}

\bibitem[{{Suleimanov} {et~al.}(2019){Suleimanov}, {Doroshenko}, \&
  {Werner}}]{suleimanov2019}
{Suleimanov}, V.~F., {Doroshenko}, V., \& {Werner}, K. 2019, \mnras, 482, 3622,
  \dodoi{10.1093/mnras/sty2952}

\bibitem[{{Tauris} \& {van den Heuvel}(2006)}]{tauris2006}
{Tauris}, T.~M., \& {van den Heuvel}, E.~P.~J. 2006, {Formation and evolution
  of compact stellar X-ray sources}, Vol.~39, 623--665

\bibitem[{{Tomsick} {et~al.}(2012){Tomsick}, {Bodaghee}, {Chaty}, {Rodriguez},
  {Rahoui}, {Halpern}, {Kalemci}, \& {{\"O}zbey Arabaci}}]{tomsick2012}
{Tomsick}, J.~A., {Bodaghee}, A., {Chaty}, S., {et~al.} 2012, \apj, 754, 145,
  \dodoi{10.1088/0004-637X/754/2/145}

\bibitem[{{Tomsick} {et~al.}(2008){Tomsick}, {Chaty}, {Rodriguez}, {Walter}, \&
  {Kaaret}}]{tomsick2008}
{Tomsick}, J.~A., {Chaty}, S., {Rodriguez}, J., {Walter}, R., \& {Kaaret}, P.
  2008, \apj, 685, 1143, \dodoi{10.1086/591040}

\bibitem[{{Tomsick} {et~al.}(2009){Tomsick}, {Chaty}, {Rodriguez}, {Walter}, \&
  {Kaaret}}]{tomsick2009b}
---. 2009, \apj, 701, 811, \dodoi{10.1088/0004-637X/701/1/811}

\bibitem[{{Tomsick} {et~al.}(2015){Tomsick}, {Krivonos}, {Rahoui}, {Ajello},
  {Rodriguez}, {Barri{\`e}re}, {Bodaghee}, \& {Chaty}}]{tomsick2015}
{Tomsick}, J.~A., {Krivonos}, R., {Rahoui}, F., {et~al.} 2015, \mnras, 449,
  597, \dodoi{10.1093/mnras/stv325}

\bibitem[{{Tomsick} {et~al.}(2016{\natexlab{a}}){Tomsick}, {Krivonos}, {Wang},
  {Bodaghee}, {Chaty}, {Rahoui}, {Rodriguez}, \& {Fornasini}}]{tomsick2016a}
{Tomsick}, J.~A., {Krivonos}, R., {Wang}, Q., {et~al.} 2016{\natexlab{a}},
  \apj, 816, 38, \dodoi{10.3847/0004-637X/816/1/38}

\bibitem[{{Tomsick} {et~al.}(2016{\natexlab{b}}){Tomsick}, {Rahoui},
  {Krivonos}, {Clavel}, {Strader}, \& {Chomiuk}}]{tomsick2016b}
{Tomsick}, J.~A., {Rahoui}, F., {Krivonos}, R., {et~al.} 2016{\natexlab{b}},
  \mnras, 460, 513, \dodoi{10.1093/mnras/stw871}

\bibitem[{{Tomsick} {et~al.}(2017){Tomsick}, {Lansbury}, {Rahoui}, {Clavel},
  {Fornasini}, {Hong}, {Aird}, {Alexander}, \& et~al.}]{tomsick2017}
{Tomsick}, J.~A., {Lansbury}, G.~B., {Rahoui}, F., {et~al.} 2017, \apjs, 230,
  25, \dodoi{10.3847/1538-4365/aa7517}

\bibitem[{{Torrej{\'o}n} {et~al.}(2010){Torrej{\'o}n}, {Schulz}, {Nowak}, \&
  {Kallman}}]{torrejo2010}
{Torrej{\'o}n}, J.~M., {Schulz}, N.~S., {Nowak}, M.~A., \& {Kallman}, T.~R.
  2010, \apj, 715, 947, \dodoi{10.1088/0004-637X/715/2/947}

\bibitem[{{Ursini} {et~al.}(2018){Ursini}, {Bassani}, {Malizia}, {Sguera},
  {Bazzano}, {Ubertini}, {Fiocchi}, \& {Bird}}]{ursini2018}
{Ursini}, F., {Bassani}, L., {Malizia}, A., {et~al.} 2018, The Astronomer's
  Telegram, 11890, 1

\bibitem[{{Winkler} {et~al.}(2003){Winkler}, {Courvoisier}, {Di Cocco},
  {Gehrels}, {Gim{\'e}nez}, {Grebenev}, {Hermsen}, {Mas-Hesse}, \&
  et~al.}]{winkler2003}
{Winkler}, C., {Courvoisier}, T.~J.~L., {Di Cocco}, G., {et~al.} 2003, \aap,
  411, L1, \dodoi{10.1051/0004-6361:20031288}

\bibitem[{{Zolotukhin} \& {Revnivtsev}(2015)}]{zolotukhin2015}
{Zolotukhin}, I.~Y., \& {Revnivtsev}, M.~G. 2015, \mnras, 446, 2418,
  \dodoi{10.1093/mnras/stu2212}

\end{thebibliography}

\label{lastpage}



\end{document}